\documentclass[aps,prd,amsmath,floats,floatfix, twocolumn,
superscriptaddress,nofootinbib,showpacs]{revtex4-1}

\usepackage[colorlinks, pdfborder={0 0 0}, plainpages=false]{hyperref}
\usepackage{graphicx}
\usepackage{xspace}
\usepackage[usenames,dvipsnames]{color}
\usepackage{amssymb}
\usepackage[dvipsnames]{xcolor}
\usepackage{placeins}
\usepackage[utf8]{inputenc}
\usepackage{lineno}
\newcommand{\D}{\mathrm{d}}
\newcommand{\lambdans}{\Lambda_\mathrm{NS}}
\newcommand{\dlambda}{\delta\lambdans^{90\%}}

\newcommand{\ii}{\mathrm{i}}
\newcommand{\chibh}{\chi_\mathrm{BH}}
\newcommand{\chins}{\chi_\mathrm{NS}}
\newcommand{\mbh}{m_\mathrm{BH}}
\newcommand{\mns}{m_\mathrm{NS}}
\newcommand{\mchirp}{\mathcal{M}_c}
\newcommand{\LL}{\mathcal{L}}

\newcommand{\arr}{\mathtt{R}}

\newcommand{\CITA}{\affiliation{Canadian Institute for Theoretical
    Astrophysics, 60 St.~George Street, University of Toronto,
    Toronto, ON M5S 3H8, Canada}} %
\newcommand{\AEI}{\affiliation{Albert Einstein Institute,
Am M\"uhlenberg, Golm, Germany}} %
\newcommand{\CIFAR}{\affiliation{Canadian Institute for Advanced Research, 
    180 Dundas St.~West, Toronto, ON M5G 1Z8, Canada}} %

\begin{document}
%\linenumbers

\title{
Measuring neutron star tidal deformability with Advanced LIGO: a Bayesian analysis
of neutron star - black hole binary observations
}

\author{Prayush Kumar}\CITA\email{prkumar@cita.utoronto.ca}
\author{Michael P\"urrer}\AEI
\author{Harald P. Pfeiffer}\CITA\CIFAR\AEI

\date{\today}

\begin{abstract}
The pioneering discovery of gravitational waves (GW) by Advanced LIGO has ushered
us into an era of observational GW astrophysics. Compact binaries remain the 
primary target sources for GW observation, of which neutron star - black hole 
(NSBH) binaries form an important subset.
GWs from NSBH sources carry signatures of (a) the tidal distortion of the
neutron star by its companion black hole during inspiral, and (b) its
potential tidal disruption near merger.
In this paper, we present a Bayesian study of the measurability of
neutron star tidal deformability $\lambdans\propto (R/M)_\mathrm{NS}^{5}$
using observation(s) of inspiral-merger GW signals from disruptive NSBH
coalescences, taking into account the crucial effect of black hole spins.
First, we find that if non-tidal templates are used to estimate source 
parameters for an NSBH signal, the bias introduced in the estimation of
non-tidal physical parameters will only be significant for loud signals with
signal-to-noise ratios greater than $\simeq30$.
For similarly loud signals, we also find that we can begin to put
interesting constraints on $\lambdans$ (factor of $1-2$) with individual
observations.
Next, we study how a population of realistic NSBH detections will improve our
measurement of neutron star tidal deformability. For an astrophysically likely
population of {\it disruptive} NSBH coalescences, we find that $20-35$ events
are sufficient to constrain $\lambdans$ within $\pm 25-50\%$, depending on the
neutron star equation of state. For these calculations we assume that LIGO will
detect black holes with masses within the astrophysical {\it mass-gap}. In case
the mass-gap remains {\it preserved} in NSBHs detected by LIGO, we estimate that
approximately $25\%$ additional detections will furnish comparable $\lambdans$
measurement accuracy.
In both cases, we find that it is the loudest $5-10$ events that provide
most of the tidal information, and not the combination of tens of low-SNR
events, thereby facilitating targeted numerical-GR follow-ups of NSBHs.
We find these results encouraging, and recommend that an effort to 
measure $\lambdans$ be planned for upcoming NSBH observations with the
LIGO-Virgo instruments.
\end{abstract}

% 04.25.D- Numerical relativity
% 04.25.dg Numerical studies of black holes and black-hole binaries
% 04.25.Nx Post-Newtonian approximation; perturbation theory; related approximations 
% 04.30.-w Gravitational waves (see also 04.80.Nn Gravitational wave detectors and experiments)
% 04.30.Db Wave generation and sources 
% 02.70.Hm Spectral methods

\maketitle

%%%%%%%%%%%%%%%%%%%%%%%%%%%%%%%%%%%%%%%%%%%%%%%%%%%%%%%%%%%%%%%%%%%%%%%%%%%%%%%
\section{Introduction}\label{s1:introduction}
%%%%%%%%%%%%%%%%%%%%%%%%%%%%%%%%%%%%%%%%%%%%%%%%%%%%%%%%%%%%%%%%%%%%%%%%%%%%%%%

The Advanced LIGO (aLIGO) observatories completed their first observing run ``O1''
early-2016, operating at a factor of $3-4$ higher gravitational-wave
(GW) strain sensitivity than their first-generation 
counterparts~\cite{Shoemaker2009}.
During O1, they made the first direct observation of gravitational 
waves~\cite{LIGOVirgo2016a}. Emitted by a pair of coalescing black holes, these
waves heralded an era of observational GW astrophysics as they traveled 
through Earth.
Towards the end of this decade, we expect aLIGO to reach its design sensitivity.
In addition to the US-based efforts, we also expect the French-Italian detector Advanced
Virgo~\cite{aVIRGO,aVirgo2}, Japanese detctor KAGRA~\cite{kagra,Somiya:2011np},
and LIGO-India~\cite{2013IJMPD..2241010U} to begin observing at comparable
sensitivities within a few years. With a global network of sensitive GW
 observatories, we can expect GW astronomy to face significant developments over the
coming years.

Coalescing compact binaries of stellar-mass black holes (BH) and/or
neutron stars (NS) are the primary targets for the second
generation GW detectors~\cite{Timmes:1995kp,Fryer:1999mi,RevModPhys.74.1015,
2010ApJ...714.1217B,2010ApJ...715L.138B,Dominik:2014yma,Belczynski:2006zi,
2012ApJ...749...91F,
Wex:1998wt,1991ApJ...379L..17N,Mandel:2015spa,Abbott:2016nhf}.
A binary system of black holes was recently observed by 
aLIGO~\cite{LIGOVirgo2016a}. Previously, stellar-mass black holes had only 
been observed by inference in mixed binaries with stellar companion (through
electromagnetic observations of the companion)~\cite{Lewin2010,
Remillard:2006fc,Fragos:2010tm}.
Neutron stars, on the other hand, have had numerous sightings. Thousands of
electromagnetically emitting neutron stars, or pulsars, have been 
documented~\cite{Manchester:2004bp},
in varied situations: as radio pulsars~\cite{Lattimer:2012nd,Manchester:2004bp},
in binary systems with a stellar companion~\cite{1971ApJ...169L..23M,
Bond:2002eh,Lattimer:2012nd,Manchester:2004bp},
and in binary neutron stars (BNS)~\cite{Hulse:1975uf,Taylor:1982wi,
Weisberg:2010zz,Lattimer:2012nd,Manchester:2004bp}.
Mixed binaries of black holes and neutron stars, is an astrophysically
interesting class of systems~\cite{Wex:1998wt,
1991ApJ...379L..17N,Janka1999,Fryer:2015jpa}, that has not yet been detected.
We expect to observe $\mathcal{O}(10)$ mixed binaries per year with
aLIGO~\cite{Abadie:2010cf}.

NSBH binaries are of interest for multiple reasons. For instance,
they have been long associated with (as possible progenitors of) short
Gamma-ray Bursts (SGRBs)~\cite{eichler:89,1992ApJ...395L..83N,moch:93,Barthelmy:2005bx,
2005Natur.437..845F,2005Natur.437..851G,Shibata:2005mz,Paschalidis2014,
Tanvir:2013}. Depending on their equation of state (EoS), NSs can get disrupted by
the tidal field of their companion BHs. Once disrupted, most of the NS
material falls into the hole over an $\mathcal{O}(1$ms$)$ time-scale,
with the rest partly getting ejected as unbound material
% (kilonovae, r-process material),
and partly forming an accretion disk around the BH.
This short lived ($0.1-1s$) disk-BH system is hypothesized to drive SGRBs
through the production of relativistic jets~\cite{Foucart:2015a,
Lovelace:2013vma,Deaton2013,Foucart2012,Shibata:2005mz,Paschalidis2014}.
However, whether or not such a system forms depends also on the nature of
the BH. Massive BHs (with $\mbh\gtrsim 12M_\odot$), as well as BHs with
large retrograde spins, tend to swallow the NS whole without forming a
disk~\cite{Foucart:2013psa}.
On the other hand, {\it low-mass} BHs with $\mbh\in[3M_\odot, 12M_\odot]$\footnote{
The upper limit on BH mass that allows for NS disruption may very well be
higher, depending strongly on the magnitude of BH spin~\cite{Foucart:2014nda}.},
can disrupt their companion NSs much before merger, forming long-sustained disks
that are required to sustain SGRBs~\cite{Shibata:2007zm,2010PhRvD..81f4026F,
Lovelace:2013vma,Foucart:2014nda,Kawaguchi:2015}.
% 
% Such systems additionally leave a strong imprint on the emitted GW signal.
A {\it coincident} detection of both GWs and gamma-rays from an NSBH merger,
will provide us with a unique opportunity to confirm this hypothesized link
between NSBH mergers and GRBs~\cite{Abbott:2016wya}.

Another question that compact object mergers can help answer is `what is the 
nature of matter at nuclear densities supported by NSs'?
A large fraction of past work aimed at measuring NS matter effects from GW
signals has consisted of inquiries about BNSs~\cite{Lee1999a,Lee1999b,Lee2000,
oechslin:07,Read:2008iy,Markakis:2010mp,Markakis:2011vd,stergioulas:11,
East:2011xa,Lackey2014,Wade:2014vqa,Bauswein:2014qla}. In this paper, we will
instead focus on NSBHs.
During the course of early inspiral, the tidal field of the BH produces a
deformation in its companion NS. The quadrupolar moment of the star associated
with this deformation also depends on its material properties, through an EoS-dependent
tidal deformability parameter $\lambdans$. This induced quadrupolar moment
changes over the orbital time-scale, resulting in the emission of GWs in {\it
coherence} with the orbital waves.
These waves draw more energy from the orbit and increase the inspiral rate (as
compared to an equivalent BBH)~\cite{Flanagan2008}.
% The coherence of GW emission between stellar and orbital waves drains energy 
% more rapidly from the orbit and increases the inspiral rate (compared to a
% BBH)~\cite{Flanagan2008}.
% We know the leading and next-to-leading order terms in post-Newtonian (PN)
% theory that capture this effect~\cite{Vines2011}, entering binary phasing at
% $5$PN order.
% 
Closer to merger, the strong tidal field of the BH can disrupt the NS. The
quadrupolar moment of the disrupted binary system falls monotonically over a
millisecond time-scale~\cite{Kyutoku:2010zd,Lackey:2013axa,Lovelace:2013vma,
Foucart:2015a,Pannarale:2015jia}, resulting in the damping of GW amplitude.
This penultimate stage also depends strongly on the internal structure and energy
transport mechanism of the NS, and carries the strongest tidal signature in the
GW spectrum~\cite{Foucart:2014nda,Deaton2013}.

% This penultimate stage carries the strongest signature in the GW spectrum, and
% can possibly be accompanied by SGRBs. 
% The final stage is a BH, whose quasi-normal
% modes may (or not) be strongly excited~\cite{FoucartEtAl:2011,Lackey:2013axa}.

Gravitational waves emitted by coalescing NSBH binaries carry subtle hints of
the NS EoS from inspiral through to merger. During early inspiral, the tidal
dephasing is relatively weak and has a frequency dependence equivalent to a
$5^{th}$ Post-Newtonian (PN) order effect~\cite{Vines2011}. Closer to merger,
a disruptive fate of the NS
can result in a strong suppression of GW emission above a cut-off 
frequency~\cite{Pannarale:2015jia}. Some past studies of tidal measurements
with NSBH binaries have used PN inspiral-only waveforms~\cite{Maselli:2013rza}.
In doing so, however, they ignore (i) the merger signal which could contain significant
information for NSBHs, and (ii) the errors due to unknown vaccum terms in PN 
waveforms, which could dominate over the tidal terms themselves~\cite{Barkett2015,
Yagi:2014}.
Some other studies that account for merger effects via the use of complete
numerical simulations~\cite{Foucart:2013psa}, are limited in the binary
parameter space they sample.
Others, that do the same through the use of phenomenological waveform
models~\cite{Lackey2011,Lackey:2013axa} use the Fisher matrix to estimate
$\lambdans$ measurement errors. Fisher matrix estimates may become
unreliable at realistic signal-to-noise ratios (SNR)~\cite{Vallisneri:2007ev},
such as those as we might expect in the upcoming observing runs of GW
detectors~\cite{Abadie:2010cf}, and we improve such studies with a
fully Bayesian treatment of the problem here.

In this paper we study the measurability of neutron star's tidal deformability
from realistic binaries of {\it low}-mass BHs and NSs by aLIGO. We also probe
how tidal effects affect the estimation of other binary parameters for the same
class of systems. This study improves upon previous work in the following ways.
First, we include tidal effects during inspiral and merger in a consistent
way, by using the waveform model of Lackey {\it et al.}~\cite{Lackey:2013axa}
(abbreviated henceforth to ``LEA'').
Second, we include the effect of black hole spin on tidal GW signals, in
addition to the effect of BH mass, tidal deformability of the NS, and the SNR.
Third, we perform a complete Bayesian analysis, instead of using the Fisher matrix
approximation.
And fourth, we explore how our measurement errors decrease as we gain information
from multiple (realistic) events.

We now outline the main questions and results discussed in this paper.
First, we probe the effect of ignoring tidal effects in
the recovery of non-tidal binary parameters, such as 
component masses and spins. This is the case for current and planned aLIGO
efforts.
To do so, we first use the enhanced-LEA (or ``LEA+'', see Sec.~\ref{s2:waveforms})
model to generate a set of realistic signals;
and then use non-tidal (BBH) waveform filters to estimate the underlying
binary masses and spins with a Markov-chain Monte Carlo. Here and throughout,
we use the zero-detuning high-power design sensitivity curve~\cite{Shoemaker2009}
to characterize the expected detector noise.
We find that, for individual events, ignoring tidal effects will affect mass
and spin-estimation only marginally; only for very loud signals (SNRs $\gtrsim 30$)
will the systematic biases be large enough to exceed the underlying
statistical uncertainty. Furthermore, detection searches can ignore tidal
effects without loss of sensitivity.

Second, we study the ability of aLIGO to constrain neutron star tidal 
deformability with a single observation of an NSBH merger. For this, we
use the same setup for signal waveforms as before, but replace the filter
template model with one that includes tidal effects from inspiral
through to merger (i.e. LEA+)~\cite{Lackey:2013axa}. For most binaries with
BH masses outside of the mass-gap $(2-5M_\odot)$~\cite{Bailyn:1997xt,
Kalogera:1996ci,Kreidberg:2012,Littenberg:2015tpa} and/or realistic signal-to-noise
ratios (SNR), we find it difficult to put better than a factor of $2$ bound
on $\lambdans$ with a single observation. As we can see from
Fig.~\ref{fig:TT_LambdaCIWidths90_0_Lambda_SNR}, it is only at SNRs 
$\rho\gtrsim 20-30$ (under otherwise favorable circumstances, such as a stiff
equation of state) that we are able
to bring this down to a $\pm 75\%$ bound on $\lambdans$. For signals louder
than $\rho =30$, we can constrain $\lambdans$ to a much more meaningful degree
(within $\pm 50\%$ of its true value).
While this is discouraging at first, we turn to ask: what if we combine
information from a population of low-SNR observations?

The EoS of matter at nuclear densities is believed to be universal among all
neutron stars. The Tolman-Oppenheimer-Volkoff equation~\cite{Tolman:1939jz,
Oppenheimer:1939ne,1934PNAS...20..169T}
would then predict that NS properties satisfy a universal relationship
between $\lambdans$ and $\mns$. As the final part of
this paper, we combine information from multiple observations of realistic NSBH
systems and perform a fully-Bayesian analysis of how our estimation of
$\lambdans$ changes as we accumulate detections. This is similar to an earlier
study~\cite{DelPozzo:13} aimed at binary neutron stars.
% % 
We restrict ourselves to a population of NSs with masses clustered very tightly
around $1.35M_\odot$ (with a negligible variance), and negligible spins. We
sample different nuclear EoSs by sampling entire populations fixing different
values for the NS tidal deformability.
% These values correspond to a hard EoS at the lower end, and a soft EoS at the upper.
For all populations, we take source locations to be uniformly distributed in
spatial volume, and source orientations to be uniform on the $2-$sphere. To
summarize, we find the following:
(a) Our median estimate for $\lambdans$ starts out prior dominated, but 
converges to within $10\%$ of the true value within $10-20$ detections.
(b) Measurement uncertainties for $\lambdans$, on the other hand, depend on
$\lambdans$ itself. We find that for hard equations of state (with 
$\lambdans\geq 1000$), $10-20$ observations are sufficient to constrain
$\lambdans$ within $\pm 50\%$. For softer equations of state, the same level
of certainty would require substantially more ($25-40$) observations.
(c) Further, if the astrophysical ``mass-gap''~\cite{Bailyn:1997xt,
Kalogera:1996ci,Kreidberg:2012,Littenberg:2015tpa} is real, we find that $20-50\%$
additional observations would be required to attain the same measurement
accuracy as above. (d) Putting tighter constraints on the $\lambdans$ of a
population would require $50+$ NSBH observations, in any scenario.
And, (e) it is the loudest $5-10$ events that will furnish the bulk
of tidal information, and not the combination of a large number of 
low-SNR events.
All of the above is possible within a few years of design
aLIGO operation~\cite{Abadie:2010cfa}.

In this paper, we restrict our parameter space to span mass-ratios
$q:=\mbh/\mns\in[2,5]$, dimensionless BH spin (aligned with orbit)
$\chibh\in[-0.5, +0.75]$, and dimensionless NS tidal deformability 
$\lambdans:= G\left(\frac{c^2}{G \mns}\right)^5\lambda \in[500, 2000]$.
These ranges are governed by the calibration of the LEA+ model which we use as
filters. 
Most of the disruptive NSBH simulations that LEA+ has been calibrated to
involve $1.35M_\odot$ NSs, and it is unclear how reliable the model is 
for different NS masses~\cite{Lackey:2013axa,Pannarale:2015jka}. This
motivates us to conservatively fix NS masses to $1.35M_\odot$ in our 
simulated signals (not templates). But, since the domain of calibration of
LEA+ excludes NS spin completely, we fix $\chins=0$ in both signals as well as
filter templates. We expect the effect of ignoring NS mass and spin variations
in our NSBH populations to be less severe than for
BNSs~\cite{Agathos:2015a}, considering the higher mass-ratios of NSBHs.
The accuracy of our quantitative results depends on the reliability of LEA+,
which is the only model of its kind in current literature. A more recent 
work~\cite{Pannarale:2015jka} improves upon the amplitude description of LEA+,
but needs to be augmented with a compatible phase model. Overall, we expect
our broad conclusions here to hold despite modeling inaccuracies (with errors
not exceeding $\mathcal{O}(10\%)$~\cite{Pannarale:2015jka}).
Finally, our results apply to LIGO instruments at design sensitivity,
which they are projected to attain by $2019$~\cite{Shoemaker2009,
Abbott:2016wya}.

The remainder of the paper is organized as follows. 
Sec.~\ref{s1:techniques} discusses data analysis techniques and resources 
used in this paper, such as the waveform model, and parameter estimation 
algorithm.
Sec.~\ref{s1:PEwithnoNS} discusses the consequences of ignoring tidal 
effects in parameter estimation waveform models.
Sec.~\ref{s1:PEwithNS} discusses the measurability for the leading order
tidal parameter $\lambdans$ at plausible SNR values.
Sec.~\ref{s1:multiple_observations} discusses the improvement in our
measurement of $\lambdans$ with successive (multiple) observations of
NSBH mergers.
Finally, in Sec.~\ref{s1:discussion} we summarize our results and discuss
future prospects with Advanced LIGO.

%%%%%%%%%%%%%%%%%%%%%%%%%%%%%%%%%%%%%%%%%%%%%%%%%%%%%%%%%%%%%%%%%%%%%%%%%%%%%%%
\section{Techniques}\label{s1:techniques}
%%%%%%%%%%%%%%%%%%%%%%%%%%%%%%%%%%%%%%%%%%%%%%%%%%%%%%%%%%%%%%%%%%%%%%%%%%%%%%%
% #################
% THESE FIGURES ARE FOR THE NEXT SECTION
\begin{figure*}
\centering 
\includegraphics[trim=20 18 18 18 0,clip=true,width=1.8\columnwidth]{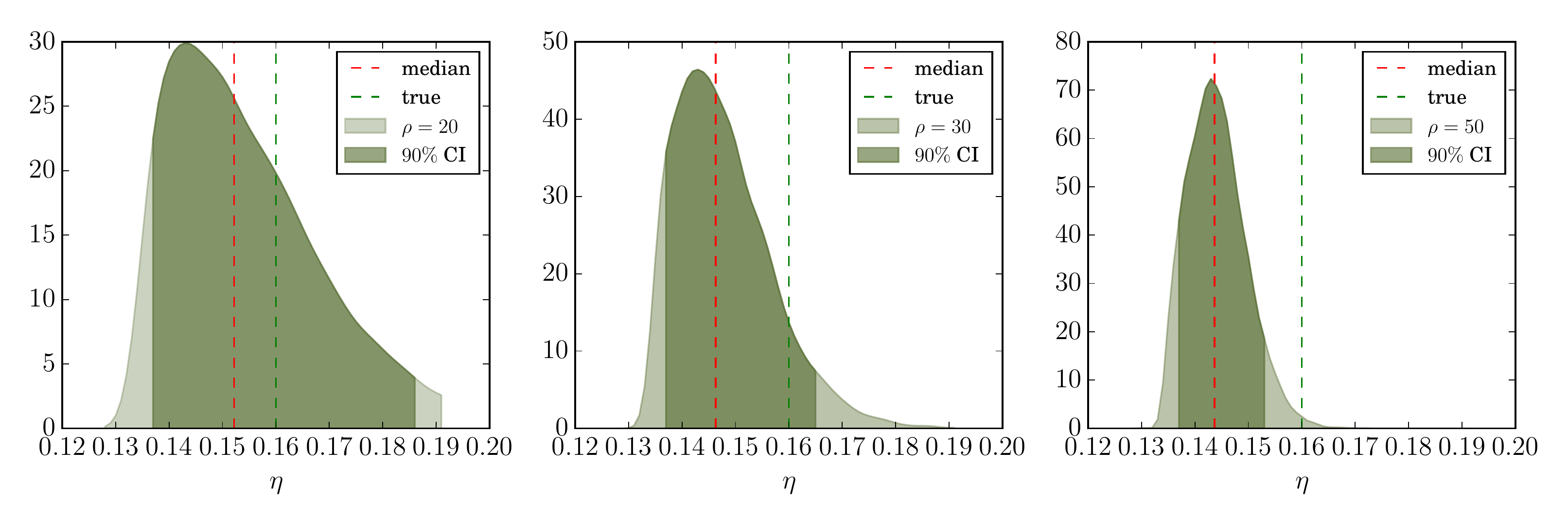}\\
\caption{{\bf Illustrative posterior probability distributions for mass-ratio $\eta$ at different SNR values:}
We show here probability distributions for mass ratio $\eta$ as measured
for the same signal at different SNRs. The intrinsic parameters of the source
are: $q = \mbh/\mns = 5.4M_\odot/1.35M_\odot = 4$, $\chibh=+0.5$, and $\lambdans=2000$;
and the signal is injected at SNRs $\rho=\{20,30,50\}$ (left to right). The templates
{\it ignore} tidal effects.
In each panel: the dashed red line marks the median value
$\eta^\mathrm{Median}$, while the dashed green line show the true value
$\eta^\mathrm{Injected}$. The darker shading shows
the recovered $90\%$ credible interval for $\eta$, $(\Delta\eta)^{90\%}$.
Comparing systematic and statistical errors, we find that:
at $\rho=20$, $\eta$ measurement is dominated by statistical
errors; at $\rho=30$, the two become comparable; and 
for louder signals ($\rho\simeq50$), the systematic errors dominate.
}
\label{fig:SingleSystemEtaPDFvsSNR}
\end{figure*}

\subsection{Waveform Models}\label{s2:waveforms}

Lackey {\it et al.}~(LEA)~\cite{Lackey:2013axa} developed a complete inspiral-merger
waveform model for disrupting NSBHs. Theirs is a frequency-domain
phenomenological model that includes the effect of BH and NS masses and spins
$\{\mbh, \chibh, \mns\}\equiv\vec{\theta}$ and NS tidal deformability
$\lambdans$. It was calibrated to a suite of $134$ numerical relativity (NR)
simulations of NSs inspiraling into spinning BHs, with
NS masses ranging between $1.2M_\odot\leq\mns\leq 1.45M_\odot$,
mass-ratios $2\leq q\leq 5$, and BH spins $-0.5\leq\chibh\leq+0.75$.
They also sample a total of $21$ two-parameter nuclear EoSs to cover the
spectrum of NS deformability.
The GW strain $\tilde{h}(f)$ per the LEA model can be written as
\begin{equation}
 \tilde{h}_\mathrm{NSBH}(f, \vec{\theta}, \lambdans) = \tilde{h}_\mathrm{BBH}(f, \vec{\theta})\,A(f, \vec{\theta}, \lambdans)\,e^{\ii \Delta\Phi(f, \vec{\theta}, \lambdans)},
\end{equation}
with NS spin $\chins=0$ identically. Here, $\tilde{h}_\mathrm{BBH}$ is
an underlying BBH waveform model. In the original LEA model,
this was taken to be the SEOBNRv1
model~\cite{Taracchini:2012} of the Effective-one-body (EOB)
family~\cite{Buonanno99}. The factor $A(\cdot)$ adjusts
the amplitude of the BBH model to match that of an NSBH merger of otherwise
identical parameters, with NS-matter effects parametrized by $\lambdans$.
During early inspiral this term is set to
unity, but is a sensitive function of $\lambdans$ close to merger. The term with
$\Delta\Phi$ corrects the waveform phasing. During inspiral,
$\Delta\Phi$ is set to the PN tidal phasing corrections,
at the leading and next-to-leading orders~\cite{Vines2011}; close to merger,
additional phenomenological terms are needed. Both $A$ and $\Delta\Phi$ are
calibrated to all $134$ available NR simulations.

In this paper we use LEA for our signal and template modeling, but switch the 
underlying BBH model to SEOBNRv2 (and refer to it as enhanced-LEA or
``LEA+'')~\cite{Taracchini:2013rva}. We using the reduced-order
frequency-domain version of SEOBNRv2, which has the additional benefit of
reducing computational cost~\cite{Purrer:2015tud}. We expect this enhancement
from LEA$\rightarrow$LEA+ to make our conclusions more robust because: (a) the 
SEOBNRv2 model is more accurate~\cite{Kumar:2015tha,Kumar:2016dhh}, and (b)
the differences between the two EOB models are caused by the
inaccuracies of SEOBNRv1 during the {\it inspiral} phase, many orbits before 
merger~\cite{Kumar:2015tha}.
Since LEA only augments inspiral phasing with PN tidal terms, our
change in the underlying BBH model does not change LEA's construction, {\it and}
increases the overall model accuracy during inspiral.
Finally, we note that we approximate the full GW signal with its dominant
$l=|m|=2$ modes, that are modeled by LEA+. For use in future LIGO science
efforts, we have implemented the LEA+ model in the LIGO Algorithms
Library~\cite{LAL}.

% % % % % % % % % % % % % % % % % % % % % % % % % % % % % % % % % % % % % % % 
% % % % % % % % % % % % % % % % % % % % % % % % % % % % % % % % % % % % % % % 
\subsection{Bayesian methods}\label{s2:bayesian}

The process of measuring systematic and statistical measurement errors
involves simulating many artificial GW signals, and inferring source binary
parameters from them using Bayesian statistics.
We start with generating a signal waveform, using the model LEA+, and injecting
it in zero noise to obtain a stretch of data $d_n$. Source intrinsic parameters
$\vec{\Theta}:=\{\mbh,\mns,\chibh,\lambdans\}$ are reconstructed from this
injected signal. Extrinsic parameters $\vec{\theta}:=\{t_c,\phi_c\}$ 
representing the time of and phase at the arrival of signal are marginalized
over numerically and analytically (respectively), while source location and
orientation parameters such as its luminosity distance, sky location, inclination
and polarization angles are absorbed into a normalization, as describe later,
and subsequently maximized over. This is justified because in this paper we
consider the {\it single-detector case}. Using Bayes' theorem, the joint inferred
probability  distribution of $\vec{\Theta}$ can be evaluated as
\begin{equation}\label{eq:postprob}
 p(\vec{\Theta} | d_n, H) = \dfrac{p(d_n|\vec{\Theta}, H)\,p(\vec{\Theta} | H)}{p(d_n|H)}.
\end{equation}
Here, $p(\vec{\Theta} | H)$ is the {\it a priori} probability of binary parameters
$\vec{\Theta}$
taking particular values, given $H$ - which denotes all our collective knowledge,
except for expectations on binary parameters that enter
our calculations explicitly. Throughout this paper, we impose priors that are
uniform in individual component masses, BH spin, and the tidal deformability of
the NS. In addition, we restrict mass-ratios to $q\geq 2$, as LEA+ is not
calibrated for $1\leq q\leq 2$. $p(d_n|\vec{\Theta}, H)$ is the {\it likelihood}
of obtaining the given stretch of data $d_n$ if we assume that a
signal parameterized by $\vec{\Theta}$ is buried in it, and is given by
\begin{equation}\label{eq:likelihood}
 p(d_n| \vec{\Theta}, H) \equiv \LL(\vec{\Theta}) = \mathcal{N}\, \mathrm{exp}[-\frac{1}{2} \langle d_n - h | d_n - h\rangle ],
\end{equation}
where $h\equiv h(\vec{\Theta})$ is a filter template with parameters 
$\vec{\Theta}$, $\langle\cdot|\cdot\rangle$ is a suitably defined
detector-noise weighted inner-product\footnote{The inner product
$\langle\cdot|\cdot\rangle$  is defined as
\begin{equation}
\langle a|b\rangle \equiv 4\,\mathrm{Re}\left[\int_0^\infty \dfrac{\tilde{a}(f) \tilde{b}(f)^*}{S_n(|f|)}\,\D f\right],
\end{equation}
where $\tilde{a}(f)$ is the Fourier transform of the finite time series $a(t)$,
and $S_n(|f|)$ is the one-sided amplitude spectrum of detector noise. In this
work, we use the zero-detuning high-power design sensitivity curve~\cite{
Shoemaker2009} for Advanced LIGO, with $15$Hz as the lower frequency cutoff.},
and $\mathcal{N}$ is the normalization constant that absorbs source distance,
orientation and sky location parameters. As in Ref.~\cite{Purrer:2015nkh} we
use a likelihood that is maximized over the template
norm, allowing us to ignore the extrinsic parameters that only enter in the
template norm through $\mathcal{N}$. As a result, we only need to sample over
$\vec{\Theta}$ (or $\vec{\Theta} - \{\lambdans\}$ in the case of non-tidal
templates).
%In addition, we also maximize over coalescence time and phase.
The denominator in Eq.~\ref{eq:postprob} is the {\it a priori} probability of finding
the particular signal in $d_n$ and we assume that each injected signal is as
likely as any other. From the joint probability distribution
$p(\vec{\Theta} | d_n, H)$ so constructed, extracting the measured probability distribution
for a single parameter (say $\alpha$) involves integrating
\begin{equation}\label{eq:marginalize}
 p(\alpha | d_n, H) = \int\D \vec{\Theta}_\alpha\, p(\vec{\Theta} | d_n, H),
\end{equation}
where $\vec{\Theta}_\alpha$ is the set of remaining parameters, i.e.
$\vec{\Theta}_\alpha:=\vec{\Theta} - \{\alpha\}$.

We use the ensemble sampler Markov-chain Monte-Carlo algorithm implemented in
the {\tt emcee} package~\cite{emcee}, to sample the probability distribution 
$p(\vec{\Theta} | d_n, H)$. We run 100 independent chains, each of which is
allowed to collect 100, 000 samples and combine samples from chains that have
a Gelman-Rubin statistic~\cite{gelman1992} close to unity. This procedure yields
about 10,000 independent samples.
One simplification we make to mitigate computational cost is to set the
frequency sampling interval to $\Delta f=0.4$~Hz, which we find to be
sufficient for robust likelihoods calculations in zero noise~\cite{Purrer:2015nkh}.
We integrate
Eq.~\ref{eq:marginalize} to obtain marginalized probability distributions
for the NS tidal deformability parameter: $p(\lambdans|d_n,H)$. We will quote
the median value of this distribution as our {\it measured} value for
$\lambdans$, and the $90\%$ credible intervals associated with the distribution
as the statistical error-bars.
% 

%%%%%%%%%%%%%%%%%%%%%%%%%%%%%%%%%%%%%%%%%%%%%%%%%%%%%%%%%%%%%%%%%%%%%%%%%%%%%%%
\section{How is PE affected if we ignore NS matter effects?}\label{s1:PEwithnoNS}
%%%%%%%%%%%%%%%%%%%%%%%%%%%%%%%%%%%%%%%%%%%%%%%%%%%%%%%%%%%%%%%%%%%%%%%%%%%%%%%
% 
\begin{figure*}
\centering 
\includegraphics[trim=20 20 18 22 0,clip=true,width=1.95\columnwidth]{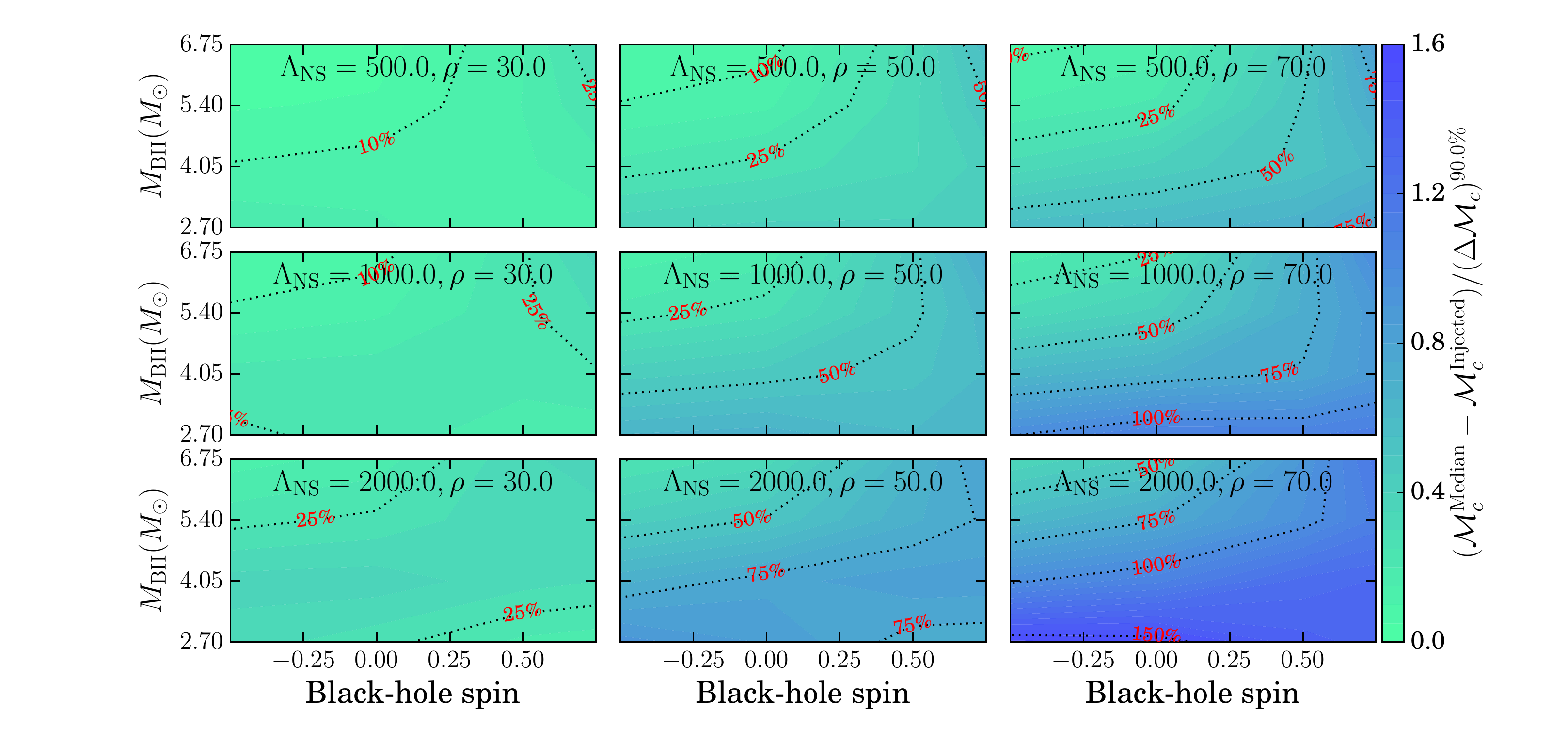}
\caption{{\bf Ratio of systematic to statistical errors in measuring $\mchirp$, ignoring tidal effects:}
We show here the ratio of systematic and statistical
measurement uncertainties for the binary chirp mass over the NSBH parameter 
space. Each panel shows the same as a function of BH mass and spin. NS mass
is fixed at $\mns=1.35M_\odot$, and its spin is set to zero. Down each column,
we can see the effect of the increasing tidal deformability of the NS at fixed
SNR. Across each row, we can see the effect of increasing the signal strength
(SNR), with the tidal deformability of the NS fixed. We show dashed contours
for $\arr_{\mchirp}=10\%, 25\%, 50\%\cdots$, with interleaving filled color
levels separated by $5\%$.
For BBHs, the statistical errors dominate systematic ones for contemporary
waveform models~\cite{Inprerp-LVC-WaveModels:2016,Kumar:2016dhh}. We find that
its not much different for NSBH binaries, until we get to very high SNRs
$\rho\gtrsim 70$.
}
\label{fig:TN_chirpMassBias_vs_Lambda_SNR}
\end{figure*}
\begin{figure*}[!t]
\centering
\includegraphics[trim=20 20 18 21 0,clip=true,width=1.95\columnwidth]{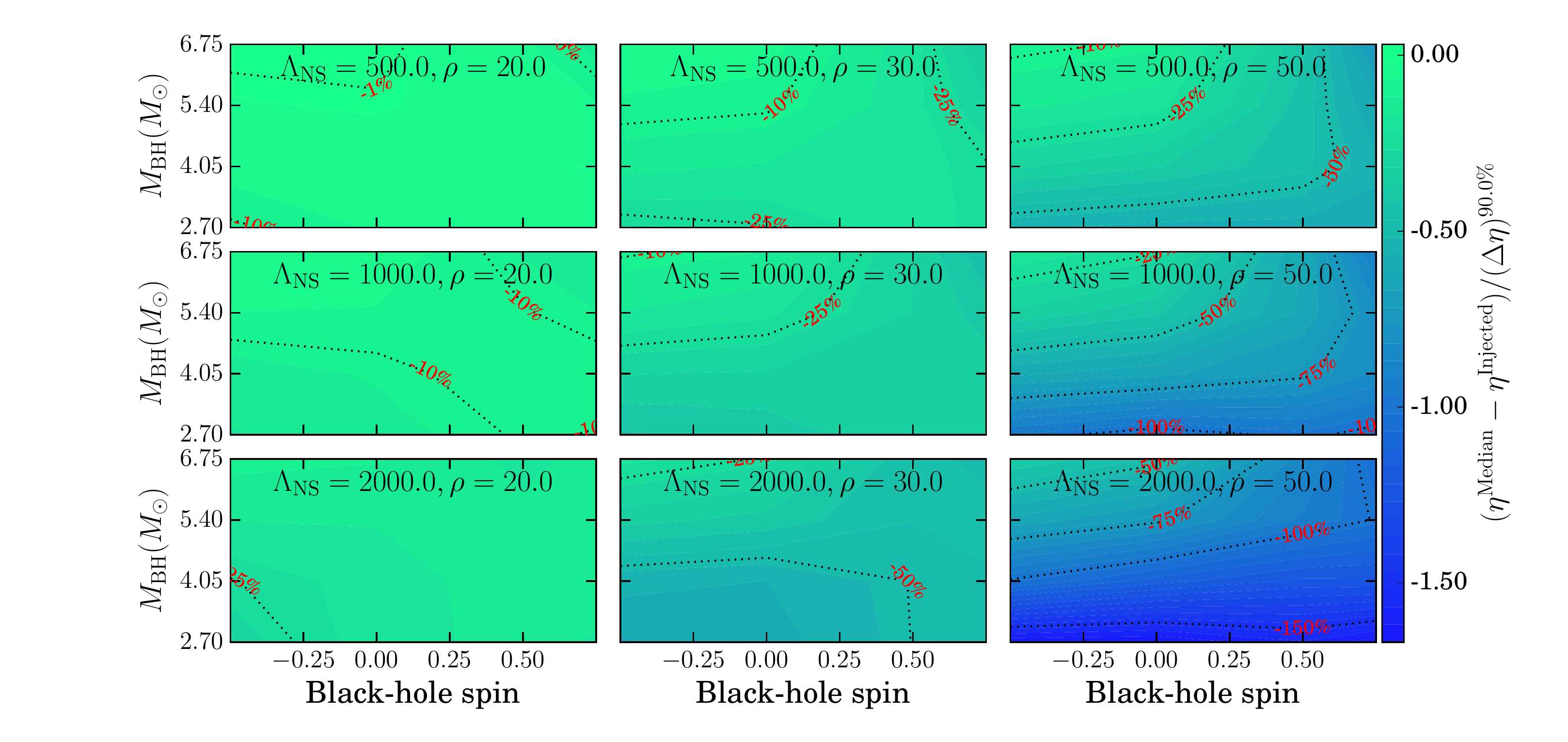}
\caption{{\bf Ratio of systematic to statistical errors in measuring $\eta$, ignoring tidal effects:}
This figure is similar to Fig.~\ref{fig:TN_chirpMassBias_vs_Lambda_SNR}
with the difference that here we show the ratio of systematic and statistical
error sources for the symmetric mass-ratio $\eta$ and not chirp mass. We find
that for fairly loud GW signals, at $\rho\simeq 50$, not including the
effects of tidal deformation of the NS on GW emission can become the dominant
source of error for astrophysical searches with Advanced LIGO. However, for
quieter signals with $\rho\leq 30$, it will have a negligible effect on the
measurement of $\eta$. We remind the reader that the SNRs here are always
single detector values.
}
\label{fig:TN_EtaBias_vs_Lambda_SNR}
\end{figure*}
\begin{figure*}
\centering
\includegraphics[trim=20 20 18 20 0,clip=true,width=1.95\columnwidth]{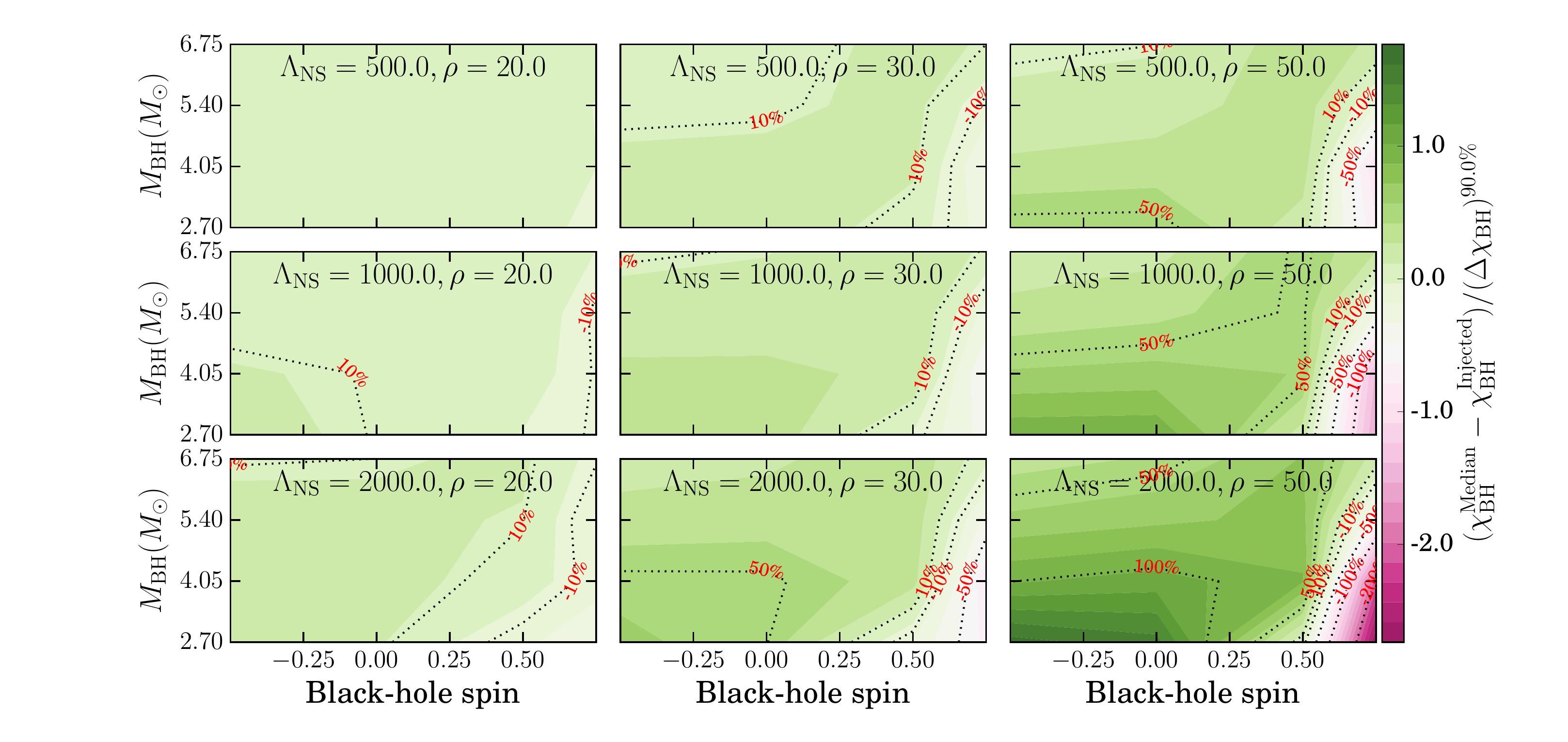}
\caption{{\bf Ratio of systematic to statistical errors in measuring $\chibh$, ignoring tidal effects:}
This figure shows the ratio of the systematic and statistical
measurement errors for BH spins $\arr_{\chibh}$. Information is arranged identically
to Fig.~\ref{fig:TN_chirpMassBias_vs_Lambda_SNR}, 
and~\ref{fig:TN_EtaBias_vs_Lambda_SNR}, with the level spacing of filled contours
increased to $15\%$.
Similar to the case of mass parameters, we find that below $\rho\approx 30$,
ignoring tidal effects in templates introduces minor systematic effects,
which remain sub-dominant to the statistical measurement uncertainties.
}
\label{fig:TN_BHspinBias_vs_Lambda_SNR}
\end{figure*}
%
% ##############################################################
% 
\begin{figure*}
\centering 
\includegraphics[trim=10 20 8 18 0,clip=true,width=1.9\columnwidth]{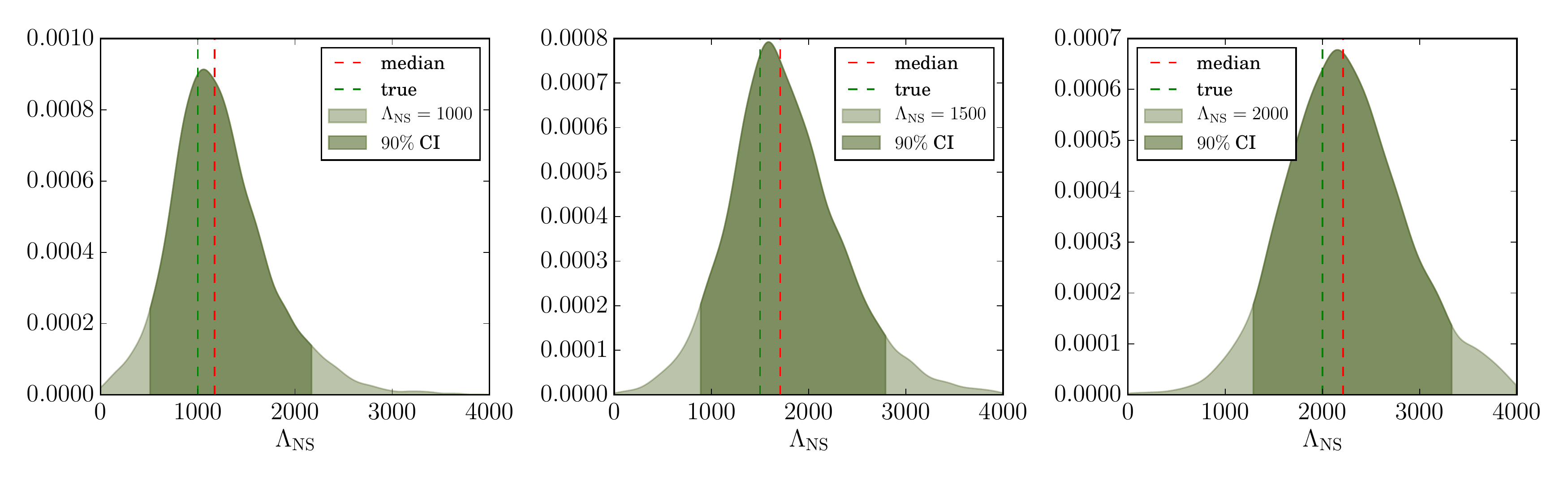}
\caption{{\bf Illustrative posterior probability distributions for NS tidal
deformability $\lambdans$:}
We show here probability distributions recovered for the NS tidal
deformability parameter $\lambdans$ from three GW injections, with parameters:
$q = \mbh/\mns = 5.4M_\odot/1.35M_\odot = 4$, $\chibh=+0.5$, and 
$\lambdans=\{1000,1500,2000\}$ from left to right. The injection SNR is fixed at
$\rho=50$. The templates {\it include} tidal effects, with a prior $0\leq\lambdans\leq 4000$.
% These figures show that this approximation holds up to (SNR)
% $\rho\simeq 30-50$.
% 
In each panel- the dashed red line marks the median value for
$\lambdans$, and the dashed green line marks its {\it true} value.
The darker shading shows the $90\%$ credible interval, whose width
$(\Delta\lambdans)^{90\%}$ is a direct measure of our statistical uncertainty.
By comparing the measurement uncertainty for these three injections, we see
that $(\Delta\lambdans)^{90\%}$ grows very slowly with $\lambdans$. Therefore,
the fractional measurement error - $(\Delta\lambdans)^{90\%}/\lambdans$ -
decreases monotonically as $\lambdans$ increases (with signal strength fixed).
}
\label{fig:SingleSystemLambdaPDFvsSNR}
\end{figure*}
Past (and future) efforts with Advanced LIGO have used (or plan to use) BBH
waveform templates to search for and characterize NSBH sources. In doing so,
they ignore the signature of NS tidal effects on the emitted GWs. In this
section we present a fully Bayesian analysis of the effect of this
simplification on the recovery of non-tidal parameters from NSBH signals.

We inject LEA+ NSBH signals into zero noise, and run an MCMC sampler on
them using equivalent BBH templates (same model, tidal terms $\rightarrow 0$).
We fix $\mns=1.35M_\odot$ and $\chins=0$, and explore a range of NS
equations of state via the single tidal deformability parameter
$\lambdans\in\{500, 800, 1000, 1500, 2000\}$. Our injections also span a
rectangular grid in the BH parameter space, with vertices at
$q\in\{2,3,4,5\}$, i.e.
$\mbh\in\{2.7M_\odot, 4.05M_\odot, 5.4M_\odot, 6.75M_\odot\}$, and BH spins
$\chibh\in\{-0.5, 0, +0.5, +0.75\}$. 
Finally, we sample all other source-related parameters, that determine the
signal strength but not character\footnote{For aligned-spin signals and
aligned-spin templates both, we only consider the contribution of the dominant
$l=|m|=2$ waveform multipoles. This approximation has the additional benefit
of combining the dependence of the waveforms on inclination, polarization
and sky location angles, as well as on distance, into the luminosity
or {\it effective} distance. This quantity only appears as an overall scaling
factor, and therefore only affects signal strength~\cite{Sathyaprakash:2009xs}.
}, by sampling the SNR $\rho\in\{20,30,50,70\}$. Our choice of injection
parameters here
is motivated by two factors: (i) previous studies of the signatures of NS tidal
effects on gravitational waves~\cite{FoucartEtAl:2011,Foucart:2013psa,
Foucart:2014nda} (which suggest that necessary conditions for the observation
of tidal effects with aLIGO include high SNRs and a low-mass spinning companion
BH); and (ii) technical constraints of our chosen LEA+ model~\cite{Lackey:2013axa}.
At design sensitivity, if we expect $0.2-300$ NSBH detections a 
year~\cite{Abadie:2010cfa}, we can expect to see $0.02-25$ 
{\it disruptive}\footnote{We assume here that BH mass values are {\it uniformly}
likely from $2M_\odot$ to $\sim 35M_\odot$~\cite{LIGOVirgo2016a}, but NSs are
disrupted in NSBH mergers only if $q\leq 6$ and $\chibh\geq 0$~\cite{Foucart:2014nda,
Foucart:2013psa}.
} NSBH mergers a year, of which we will have $0.005-7$ observations with 
$\rho\geq 20$, and $0.002-3$ a year with $\rho\geq 30$.
Therefore, our injection parameters span a physically interesting subset of NSBH
binaries, that is {\it also} likely observable in the near future. 
For our Bayesian priors, we choose uniform distributions for both component
masses and black hole spin: $\mbh\in[1.2,25]M_\odot$; $\mns\in[1.2,3]M_\odot$;
and $-0.75\leq \chibh\leq +0.75$.

The effect of ignoring
tidal corrections in templates will manifest as a systematic shift
of recovered median parameter values away from what they would be if we had used
tidal templates with identical priors. 
In zero noise, we expect the probability distributions recovered using tidal
templates to be multi-dimensional Gaussians with the maximum likelihood
parameter values approaching their true values. If the priors are not
restrictive, we expect the recovered median to also converge to the true value.
However,
the LEA model imposes significantly more restrictive priors (both mass-ratio
and spin) than SEOBNRv2~\cite{Taracchini:2013rva,Lackey:2013axa}, which shifts
the median value of parameters recovered using {\it our} tidal templates away
from their true value. If we use LEA+ priors for our non-tidal templates, it
would add a caveat to our original question 'can we estimate non-tidal NSBH
parameters with equivalent BBH templates'. Instead, we approximate the median
tidally recovered parameters by their true injected values, as one would expect
to recover with an ideal model for tidally disruptive NSBH mergers. With this
caveat, we estimate {\it systematic} measurement bias/errors as the
differences between median and {\it injected} parameter values. 
As an illustration, in Fig.~\ref{fig:SingleSystemEtaPDFvsSNR} we show the
recovered probability distributions for binary mass ratio $\eta$ for
three NSBH injections, with $\rho=20$ (left), $30$ (middle), and $\rho=50$
(right), and other parameters held fixed ($\mns=1.35M_\odot$, $\chins=0$,
$\mbh=5.4M_\odot$, $\chibh=+0.5$ and $\lambdans=2000$). In each panel, both
the {\it true} and median values of $\eta$ are marked, and we use
the shift between the red and green vertical lines as our estimate of systematic
measurement errors. Darker shading in all panels marks $90\%$ credible intervals,
whose width $(\Delta\eta)^{90.0\%}$ we use as a direct measure of our
{\it statistical} measurement uncertainty/error\footnote{We generalize the
notation $(\Delta X)^{90.0\%}$ to mean the $90\%$ credible interval width
for any measured source parameter $X$.}.
For the illustrated binary,
we see clearly that even when the signal is moderately loud, with $\rho=20$,
statistical errors dominate over systematics for $\eta$. As we turn up the SNR
further, the two error sources become comparable at $\rho\sim 30$,
and systematic errors dominate finally when $\rho\simeq 50$.

Credible intervals $(\Delta X)^{90\%}$ showing the precision with which
$X=\{\mchirp,\eta,\chibh\}$ can be measured, are presented in
Appendix~\ref{as1:nontidalerrors}.
%}
We remind ourselves that this {\it precision} is only meaningful so long as the
measurement is {\it accurate} to begin with. Therefore, we define $\arr_X$ as
the ratio between systematic and statistical errors associated with the
measurement of parameter $X$,
\begin{equation}\label{eq:arr}
\arr_X = \dfrac{(X^\mathrm{Median} - X^\mathrm{Injected})}{(\Delta X)^{90\%}},
\end{equation}
in order to compare the relative magnitude of both. Only when
$|\arr_X| \ll 1$ can we ignore tidal effects in our templates
without hampering the measurement of non-tidal parameters from NSBH signals.
When $\arr_X$ approaches a few tens of percent of unity, we can begin to favor
tidal templates for NSBH studies.

We start with calculating $\arr_{\mchirp}$ as a function of various source
parameters and show it in Fig.~\ref{fig:TN_chirpMassBias_vs_Lambda_SNR}.
$\mchirp$ is the leading order
mass combination that affects the GW strain emitted by compact binaries as they
spiral in, and is therefore determined the most precisely. We notice
immediately that for $\rho\leq 30$ the systematics are well under control
and we can obtain reliable chirp mass estimates for NSBH signals using BBH
templates.
For louder and less likely SNRs ($\rho\simeq 50$), we find that
$\arr_{\mchirp}$ can become comparable to unity, but only if the BH has
prograde spin $\chibh\gtrsim 0.4$, {\it and} the true NS tidal deformability
is large enough, s.t. $\lambdans \gtrsim 1000$.
We therefore conclude that only for very loud signals, with
$\rho\gtrsim 50-70$, will the inclusion of tidal terms in template
models improve $\mchirp$ estimation. For lower SNRs, inclusion of new
physical content in templates will instead get washed out by detector noise.
In addition, we also note that $\arr_{\mchirp}\geq 0$ always,
i.e. $\mchirp$ is always being over-estimated. This is to be expected since
the tidal deformation of the NS drains energy faster from the orbit during
inspiral (as compared to the BBH case), and its disruption close to merger
reduces GW signal power at high frequencies. Both of these effects make the
resulting signal resemble a BBH signal of higher chirp (or total) mass, although
we expect the latter effect to be dominant~\cite{Pannarale:2011pk}.

Next, in Fig.~\ref{fig:TN_EtaBias_vs_Lambda_SNR}, we show the ratio of
measurement errors $\arr_\eta$ for the symmetric mass-ratio.
Going through the figure from left to right, we find that for realistic
SNRs ($\rho\leq 30$) the systematics remain below statistical errors for
$\eta$ measurement. The worst case is of the most deformable NSs
($\lambdans = 2000$), but even for them systematics in $\eta$ are $2\times$
smaller than the statistical measurement errors.
Moving to louder signals with $\rho\simeq 50$, we find that for binaries of
fairly deformable NSs ($\lambdans\gtrsim 1500$) and low-mass BHs
($\mbh\leq 5M_\odot$) that have prograde spins ($\chibh\gtrsim +0.4$), our
measurement of mass-ratio can be seriously compromised by ignoring tidal
physics in template models. 
This pattern is continued at even higher SNRs, as we can see from
Fig.~\ref{fig:TN_EtaBias_vs_Lambda_SNR}. We therefore conclude that, even if
under moderate restrictions on BH and NS parameters, $\rho=30-50$ is loud enough
to motivate the use of tidal templates in aLIGO data analyses.
In addition, we also notice that, unlike for $\mchirp$, the median value of
$\eta$ is always {\it lower} than its true value, which is what we expect if we
want BBH templates to fit NSBHs that disrupt and merge at lower frequencies.

Moving on from mass to spin parameters, we now consider the measurement of BH
spin angular momentum $\chibh$. The ratio of systematic and statistical errors
for $\chibh$ are shown in
Fig.~\ref{fig:TN_BHspinBias_vs_Lambda_SNR}. The presentation of information in this 
figure is identical to that of Fig.~\ref{fig:TN_chirpMassBias_vs_Lambda_SNR}
and~\ref{fig:TN_EtaBias_vs_Lambda_SNR}. A diverging colormap is used so that both 
extremes of the colorbar range point to large systematic biases, while its zero (or
small) value lies in the middle.
For the lowest SNR considered ($\rho=30$), $\chibh$ bias is about $2\times$
smaller than its statistical measurement uncertainty, and is therefore mostly
negligible. Both do become somewhat comparable, but only when we have the most
deformable NSs in orbit around low-mass BHs. 
At higher SNRs $(\rho\simeq50-70)$, we find that the systematics in $\chibh$
measurement can dominate completely, especially for binaries containing
mass-gap violating BHs and/or deformable NSs with $\lambdans\geq1000$.
From Fig.~\ref{fig:TN_BHspinBias_vs_Lambda_SNR} we additionally note that when
the source spin magnitudes approach the highest allowed, i.e. at both extremes
of the $x$-axes, $\chibh\times\arr_{\chibh}<0$. This is to be expected because
the median of the recovered posterior distributions for $\chibh$ can only get
pushed inwards from the boundaries.

Summarizing these results, we find that irrespective of system parameters,
below a signal-to-noise ratio of $30$, our measurements of mass and spin
parameters of astrophysical NSBH binaries will remain limited by the intrinsic
uncertainty due to instrument noise, and do not depend on whether we include
tidal effects in template models. However, when the signal-to-noise ratio
exceeds $30$ the systematic bias in binary mass and spin measurements become
comparable to and can exceed the uncertainty due to noise. Of the different
non-tidal parameters considered, we find that the measurement of $\eta$
degrades worst (in a relative-error sense) due to the use of BBH templates in
deciphering an NSBH signal. Of all the sub-categories, we find that tidal
templates could especially help with the parameter estimation of astrophysical
{\it mass-gap violating} NSBH binaries,

%%%%%%%%%%%%%%%%%%%%%%%%%%%%%%%%%%%%%%%%%%%%%%%%%%%%%%%%%%%%%%%%%%%%%%%%%%%%%%%
\section{What do we gain by using templates that include NS matter effects?}\label{s1:PEwithNS}
%%%%%%%%%%%%%%%%%%%%%%%%%%%%%%%%%%%%%%%%%%%%%%%%%%%%%%%%%%%%%%%%%%%%%%%%%%%%%%%
% % 
% #################
\begin{figure*}
\centering    
\includegraphics[trim=35 21 15 21 0,clip=true,width=2.2\columnwidth]{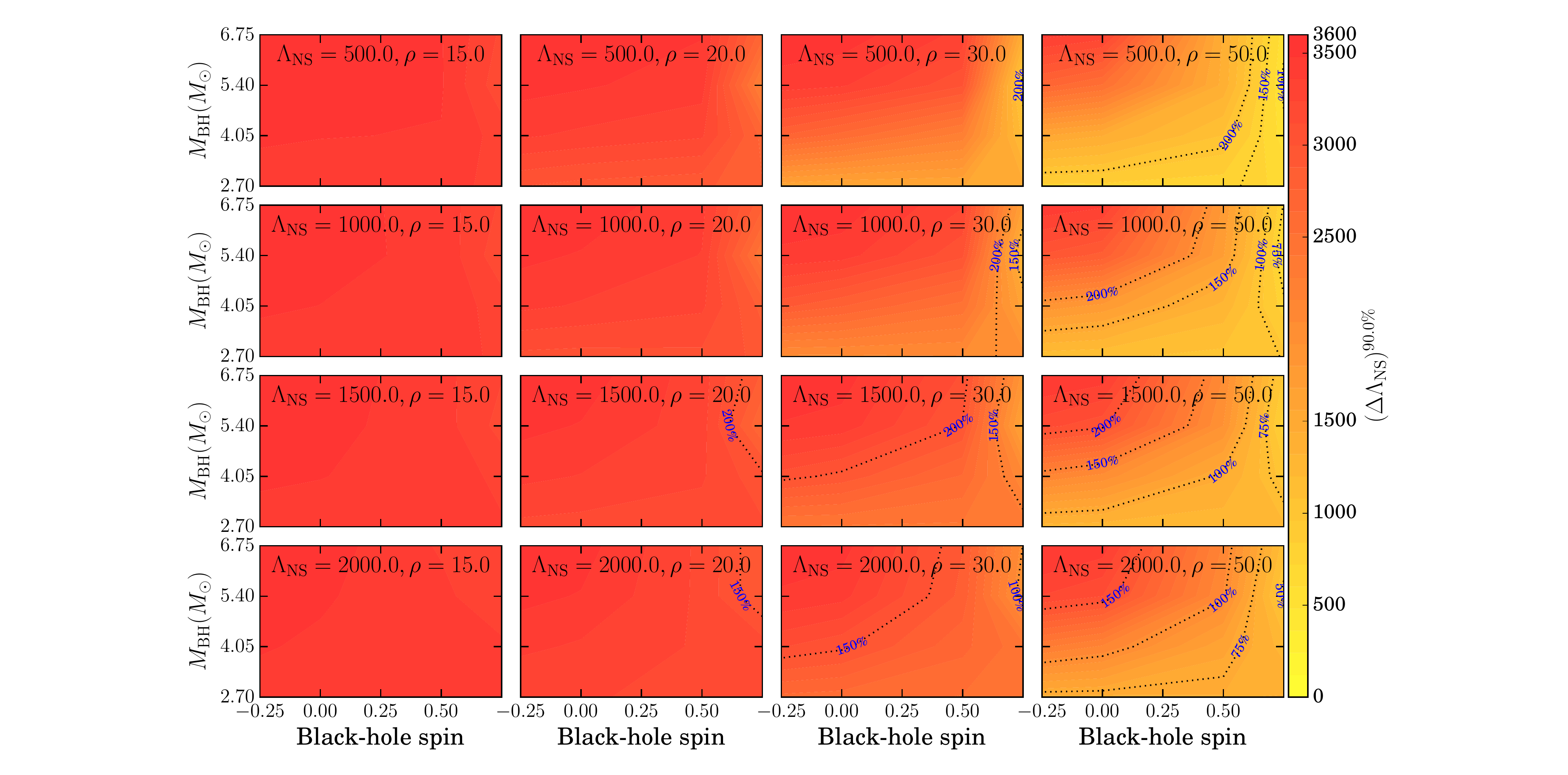}
\caption{{\bf Statistical uncertainty in $\lambdans$ measurement:}
Here we show the statistical uncertainty in the measurement of
$\lambdans$. In each panel, the
same is shown as a function of the BH mass and spin, keeping $\lambdans$ and
injection's SNR $\rho$ fixed (noted in the panel). Rows contain panels
with the same value of $\lambdans$, with $\rho$ increasing from left to right.
Columns contain panels with the same value of $\rho$, with $\lambdans$ 
increasing from top to bottom.
Contours at $(\Delta\lambdans)^{90\%}=\{50\%, 75\%, 100\%, 150\%, 200\%\}\times\lambdans^\mathrm{Injected}$ demarcate regions where we can constrain the
$\lambdans$ parameter well (within a factor of two of the injected value).
We note that, as expected, the measurement accuracy for $\lambdans$ improves
with (i) increasing SNR, (ii) increasing $\lambdans$, (iii) increasing BH spin,
and (iv) decreasing BH mass.
}
\label{fig:TT_LambdaCIWidths90_0_Lambda_SNR}
\end{figure*}
% % 
\begin{figure}
\centering    
\includegraphics[trim=10 10 0 10 0,clip=true,width=1.05\columnwidth]{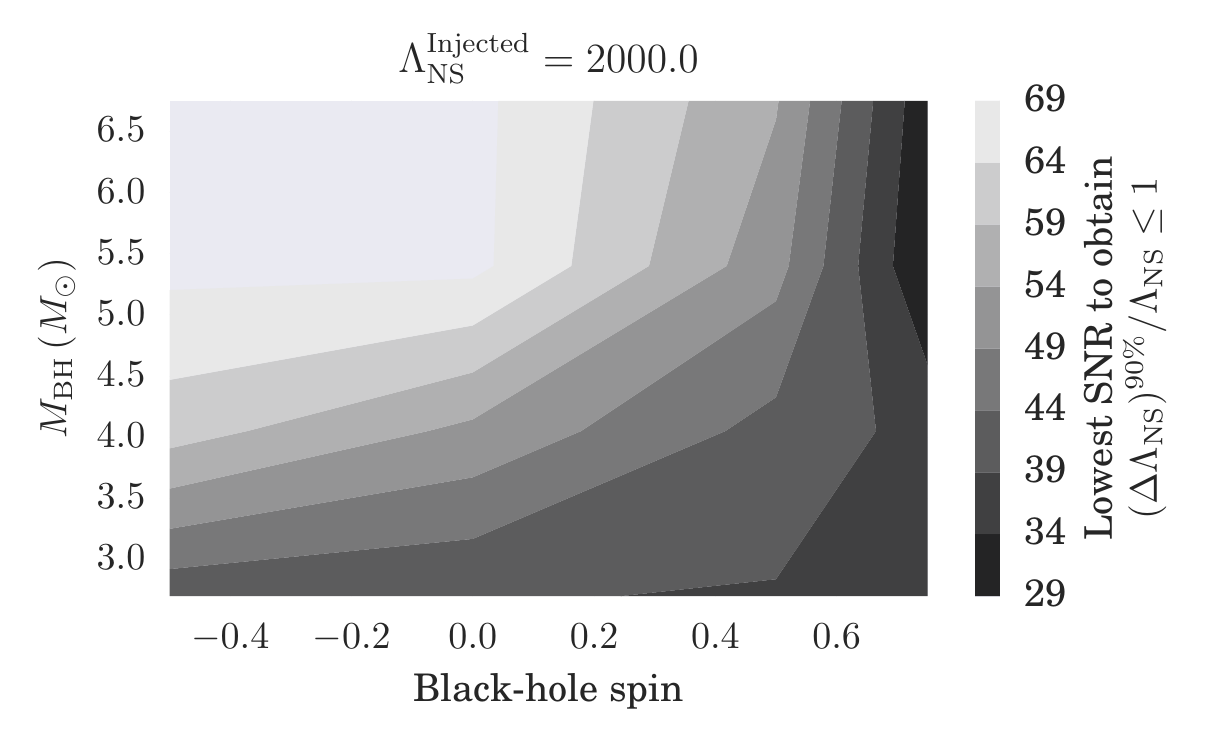}
\caption{
We show here, as a function of BH mass and spin, the {\it minimum} signal
strength (SNR) required to constrain $\lambdans$ within an interval of width
equal to $100\%$ of its true value, i.e. with $\pm 50\%$ error-bars. The NS mass
is fixed at $1.35M_\odot$, spin at zero, and $\lambdans=2000$.
We can see that, even in the most conducive circumstances with large aligned 
$\chibh$ and a comparable mass BH, we can only constrain $\lambdans$ to better
than $\pm 50\%$ {\it if} the SNR is $\gtrsim 29$. In the era of design
sensitivity LIGO instruments, we expect this to happen approximately once in a
year of observation~\cite{Abadie:2010cfa}.
}
\label{fig:TT_SNRThresholds_BHspin_BHmass_CI90_0}
\end{figure}
% %

%

In the previous section, we showed that the effects of the tidal deformation of
NSs by their companion BHs become discernible in
the GW spectrum under certain favorable conditions, including (a) BH mass is
sufficiently small, (b) BH spin is positive aligned, i.e. $\chibh\gtrsim +0.4$,
(c) the NS is not very compact, with $\lambdans\gtrsim 1000$, and (d) the
source location and orientation are such that its GW SNR $\gtrsim 30$.
Both condition (a) and (b) enhance the tidal distortion of the star and increase
the number of orbits the system goes through at small separation, where the
differences between NSBH and BBH signals are maximal.
Conditions (a)-(c) also reduce the onset frequency of the disruption of the NS,
allowing for it to happen earlier in the orbit. 
We expect that these conditions are also the ones which should maximize the
likelihood of {\it measuring} tidal effects in NSBH signals. Here,
we turn the question around to ask: under similarly favorable circumstances,
can we gain insights about the internal structure of neutron stars from GW
observations?

In this section, we calculate the accuracy with which we measure $\lambdans$ from
{\it single} GW observations. We sample the same set of disruptive NSBH mergers
as in the previous section, i.e. those with $q=\{2,3,4,5\}$,
$\chibh=\{-0.5,0,+0.5,+0.75\}$, and $\lambdans=\{500, 800, 1000, 1500, 2000\}$;
fixing the NS mass $\mns=1.35 M_\odot$ and $\chins=0$. For each unique
combination of these parameters, we inject LEA+ signals into zero noise and
perform a fully Bayesian parameter estimation analysis of each with LEA+
templates. Our  priors on component masses and spins remain as in the
previous section, with mass-ratio additionally restricted to $2\leq q\leq 6$,
and $\lambdans$ sampled uniformly from $[0,4000]$.
As an illustration of individual injections, we show the recovered probability
distribution for $\lambdans$ for three specific configurations in 
Fig.~\ref{fig:SingleSystemLambdaPDFvsSNR}. We fix
$q = \mbh /\mns = 5.4M_\odot/1.35M_\odot = 4$, with $\chibh=+0.5$, and
vary $\lambdans$ over $\{1000, 1500, 2000\}$ between the three panels.
The SNR is fixed at $\rho=50$. The darker shaded regions mark the $90\%$ credible
interval on $\lambdans$. We note that $\lambdans$ is estimated to within
$\pm 2000$ of its true value at this SNR. Another interesting thing to note 
is that while $(\Delta\lambdans)^{90\%}$ slowly grows with $\lambdans$, the
fractional uncertainty
\begin{equation}
\delta\lambdans^{90\%}:= (\Delta\lambdans)^{90\%}/\lambdans
\end{equation}
decreases instead.
%
% \textcolor{blue}{%
Further illustrations, showing the correlation between tidal and non-tidal
parameters, are presented in Appendix~\ref{as1:illustrations}.
We will continue here to focus on the measurement of $\lambdans$ itself.

In Fig.~\ref{fig:TT_LambdaCIWidths90_0_Lambda_SNR} we show the main results of
this section. In each panel, as a function of black hole mass and spin, we show
the measured $90\%$ credible interval widths $(\Delta\lambdans)^{90\%}$. These
correspond to the full width of the dark shaded regions in the illustrative
Fig.~\ref{fig:SingleSystemLambdaPDFvsSNR}. The effect of increasing signal
strength can be seen as we go from left to right in each row. The effect of the
NS tidal deformability parameter $\lambdans$ on its own measurability can be
seen by comparing panels within each column, with the NS becoming more
deformable from top to bottom. 
A uniform pattern emerges in the left-most column, which corresponds to $\rho=20$.
We find that at this signal strength, our measurement of $\lambdans$ is
dominated by the width of our prior on it. The $90\%$ credible intervals span
the entire allowed range for $\lambdans$, making a reasonable estimation of
$\lambdans$ at $\rho\simeq20$ difficult.
Increasing the signal strength to $\rho=30$ gives marginally better results,
bringing down the statistical uncertainties to within $\pm 75-100\%$ of the
true $\lambdans$ value~\footnote{The symmetric error-bars of $\pm\mathrm{X}\%$
correspond to $\dlambda = 2\mathrm{X}\%$.}.
It is not until we reach an SNR as high as $\rho\gtrsim 50$, can we put
meaningful (i.e. $\mathcal{O}(10\%)$) constraints on $\lambdans$. For e.g.,
with a {\it single} observation of a $q=4$ binary with $\chibh\geq 0.6$ and
$\rho = 50$~\footnote{For an optimally oriented source with
$q=4, \mns=1.35M_\odot, \chibh=0.6$, an SNR of $\rho = 50$ corresponds to
a luminosity distance of $\approx 113$Mpc.}, we would be able to estimate 
$\lambdans$ to within $\pm 40\%$ of its true value (which is equivalent to
measuring the ratio of NS radius to mass with an uncertainty of about
$\pm 10\%$).
These results agree well with Sec.~\ref{s1:PEwithnoNS}, and are consistent with
Fisher matrix estimates at high SNRs~\cite{Lackey:2013axa}.

Amongst other source parameters, BH mass and spin play a dominant role. A smaller
BH with a larger spin always allows for a more precise measurement on $\lambdans$.
We can see this in the bottom right corner of each panel in
Fig.~\ref{fig:TT_LambdaCIWidths90_0_Lambda_SNR}, which corresponds to low-mass BHs
with large spins, and is simultaneously the region of smallest measurement errors on $\lambdans$.
The actual deformability of the NS also plays an important role on its own
measurability. For e.g., when $\lambdans\leq 1000$, it is fairly difficult
to meaningfully constrain $\lambdans$ without requiring the source to be
close ($\approx 100$Mpc) with a GW SNR $\rho\gtrsim 50$. Quantifying this further,
in Fig.~\ref{fig:TT_SNRThresholds_BHspin_BHmass_CI90_0} we show the minimum
signal strength required to attain a certain level of credibility in our
$\lambdans$ measurement, as a function of BH properties. The NS is allowed
the most favorable (hardest) EoS considered, with $\lambdans^\mathrm{true}=2000$.
We first note that, even with the most favorable BH and NS properties, achieving
a $\pm 50\%$ measurement certainty on $\lambdans$ will require a GW SNR
$\rho\gtrsim 30$. If we additionally restrict BH masses to lie outside of the so-called
astrophysical mass-gap~\cite{Bailyn:1997xt,Kalogera:1996ci,Kreidberg:2012,
Littenberg:2015tpa}, we will simultaneously need to restrict BH spins
to $\chibh\gtrsim +0.5$ to obtain the same measurement credibility at the same
source location.

It is interesting to note that the parameter ranges most favorable to
the measurability of $\lambdans$ are also those which produce
relatively more massive post-merger disks~\cite{Foucart2012}. That is, 
the subset of NSBHs that potentially produce SGRBs (using a sufficiently-large
disk mass as an indicator) would be the same subset most favorable
for measurement of tidal effects. Therefore the rate of SGRBs in the 
local universe (allowing for the fraction that are produced by NSBHs versus
BNSs) would be an indicator of the rate of events most favorable for nuclear
equation of state measurements.

In summary, with a single moderately loud ($\rho\lesssim 30$) GW signal from
a disruptive BHNS coalescence, we can constrain
the NS compactness parameter $\lambdans$ within $\pm 100\%$ of its true value.
To measure better with one observation, we will need a more fine-tuned source, with
$\rho\geq 30$ and high BH spins, or $\rho\geq 50$.
Finally, we note that these results are {\it conservative}, and 
BHs with spins $\chibh > 0.75$ will prove to be even more favorable laboratories
for $\lambdans$ measurement. However, we are presently unable to explore this case
in quantitative detail due to waveform model restrictions~\cite{Lackey:2013axa},
which will also restrict our analyses of GW signals during the upcoming LIGO
observing runs.

%%%%%%%%%%%%%%%%%%%%%%%%%%%%%%%%%%%%%%%%%%%%%%%%%%%%%%%%%%%%%%%%%%%%%%%%%%%%%%%
\section{Combining observations: looking forward with Advanced LIGO}\label{s1:multiple_observations}
%%%%%%%%%%%%%%%%%%%%%%%%%%%%%%%%%%%%%%%%%%%%%%%%%%%%%%%%%%%%%%%%%%%%%%%%%%%%%%%
% 
\begin{figure}
\centering    
\includegraphics[trim=18 18 18 10 0,clip=true,width=\columnwidth]{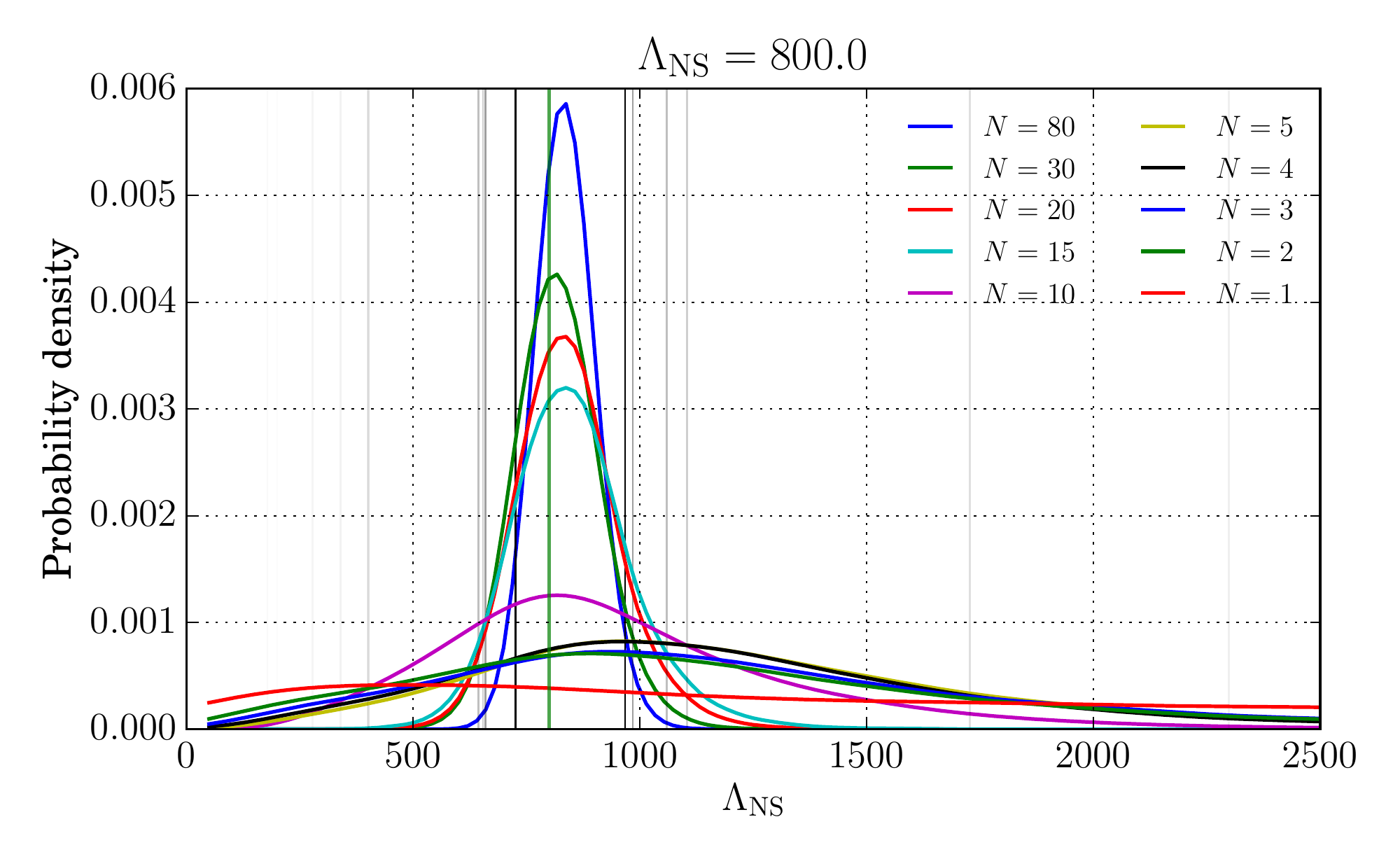}\\
\includegraphics[trim=18 18 18 10 0,clip=true,width=\columnwidth]{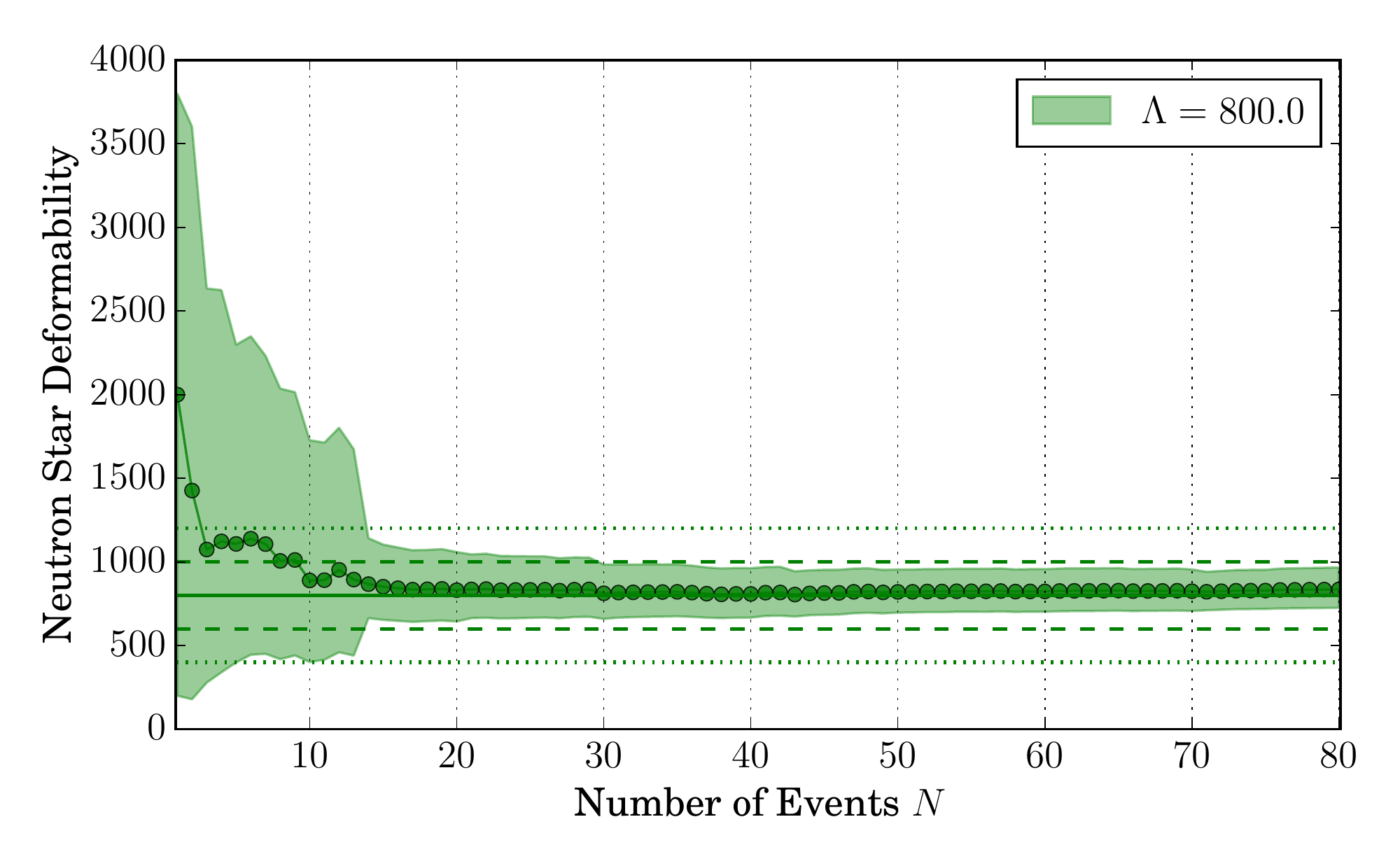}
\caption{{\bf Recovery of $\lambdans$ for an increasingly large population of BH-NS signals.}
{\it Top}: Posterior probability distributions for $\lambdans$ (colored curves), and
associated $90\%$ credible intervals (grey vertical lines), shown for different number
of accumulated observations N. Distributions are normalized to unit area.  
{\it Bottom}: Measured median value of $\lambdans$ (as solid circles) and the
associated $90\%$ credible intervals (as the vertical extent of filled region), shown as
a function of number of observations N. Solid horizontal line indicates the true value of
$\lambdans=800$. Dashed and dotted horizontal lines (a pair for each line-style) demarcate
$\pm 25\%$ and $\pm 50\%$ error bounds.
}
\label{fig:TT_Lambda_vs_N_L800_CI90_0}
\end{figure}
\begin{figure}
\centering    
\includegraphics[width=1.05\columnwidth,trim=1cm 0 0 0]{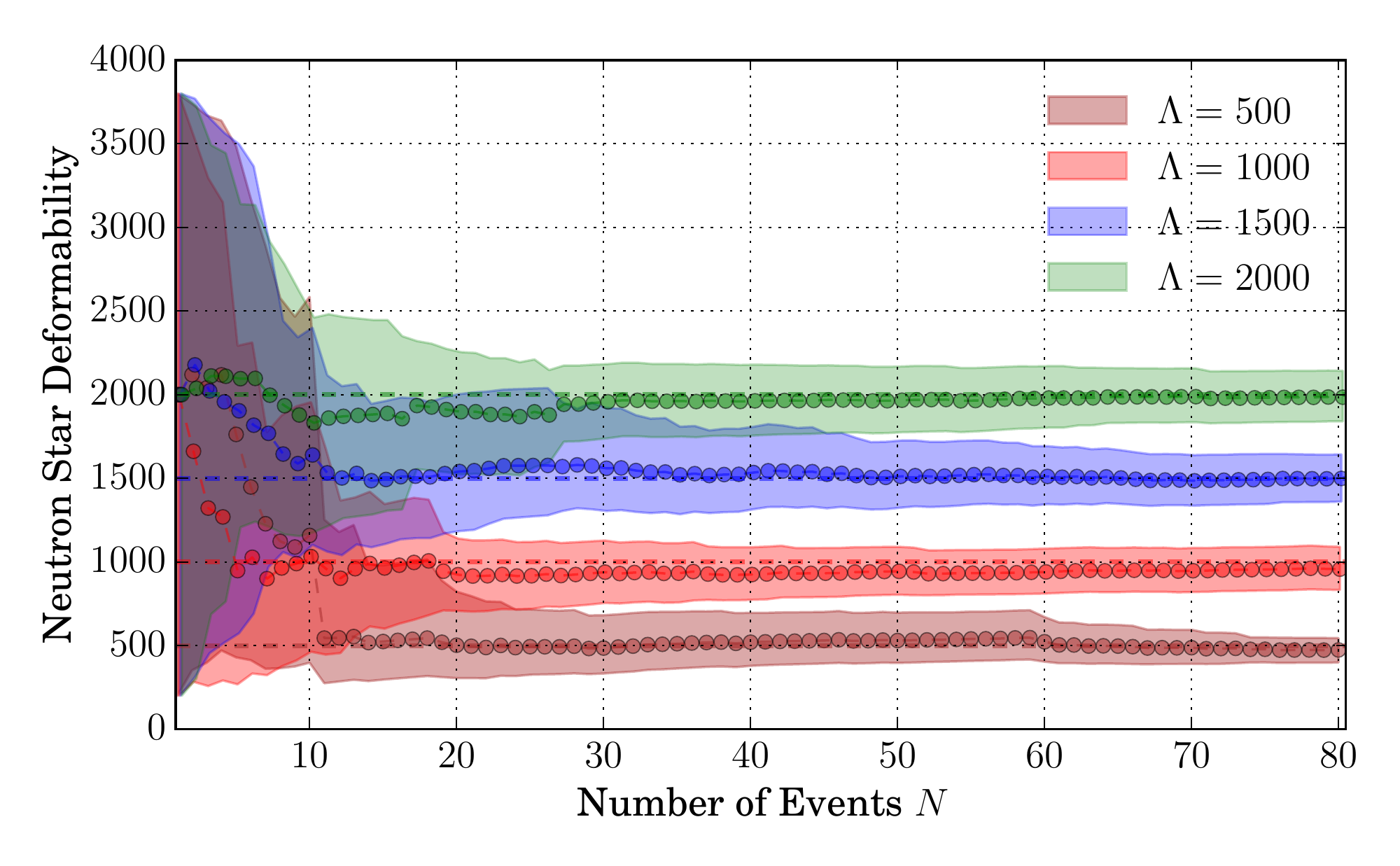}
\caption{{\bf Improvement in $\lambdans$ measurement accuracy for different NS EoS:}
In this figure, the filled regions show how our measurement of $\lambdans$
improves as the number of observed events ($N$, shown on $x$-axis) increases.
Each color corresponds to an independent population with its true value of
$\lambdans$ given in the legend. For each population, we show the median 
$\lambdans$ value (as filled circles), as well as the associated
$90\%$ credible intervals for the measurement (as the vertical extent of the
filled region about the median), as functions of $N$.
}
\label{fig:TT_Lambda_vs_N_CI90_0}
\end{figure}
\begin{figure*}
\centering
% \textbf{No Mass-Gap \hspace{6cm} Mass-Gap}\par\medskip
\includegraphics[trim=1cm 0 0 0, width=1.025\columnwidth]{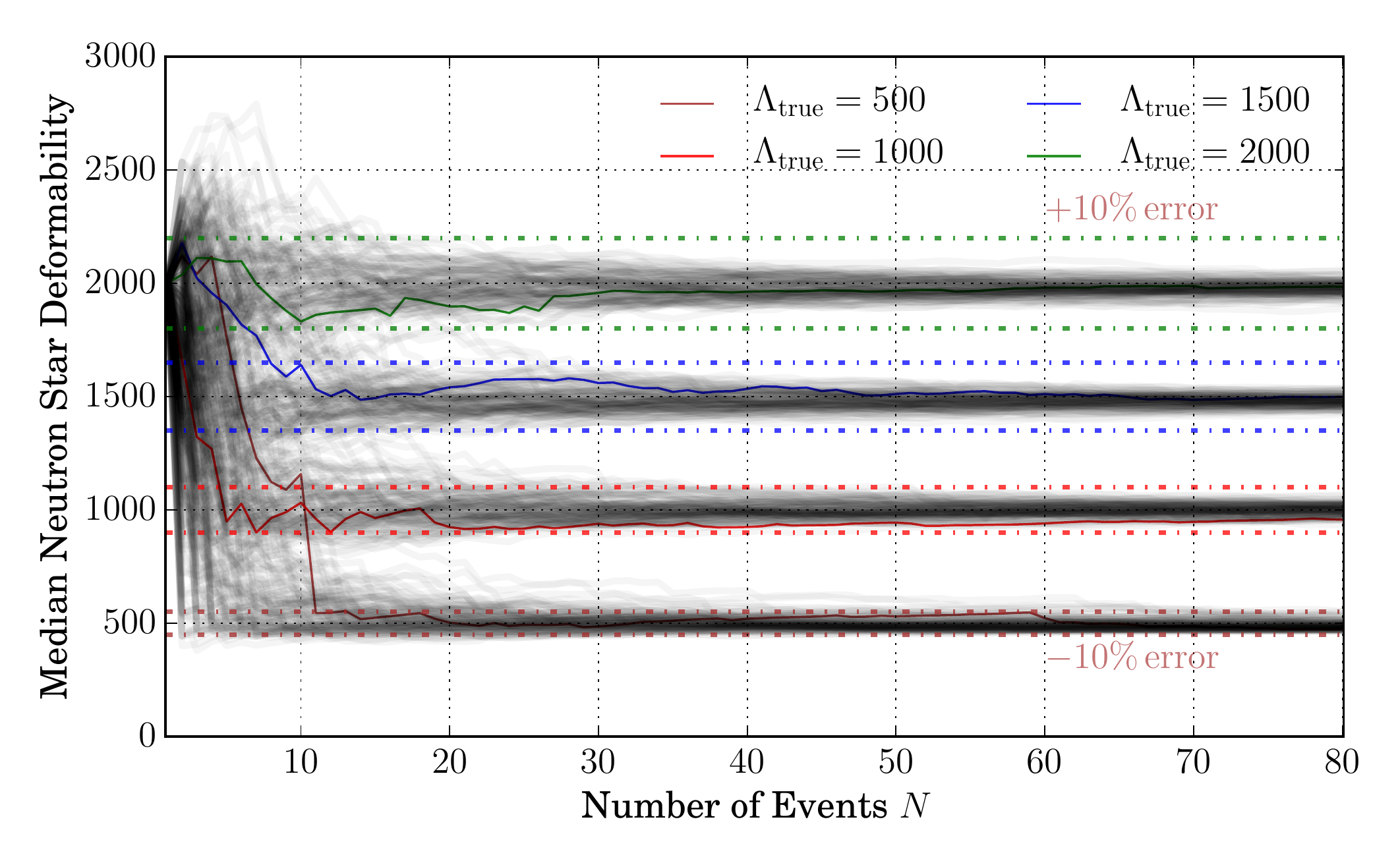}
\includegraphics[trim=0 0 1cm 0, width=1.025\columnwidth]{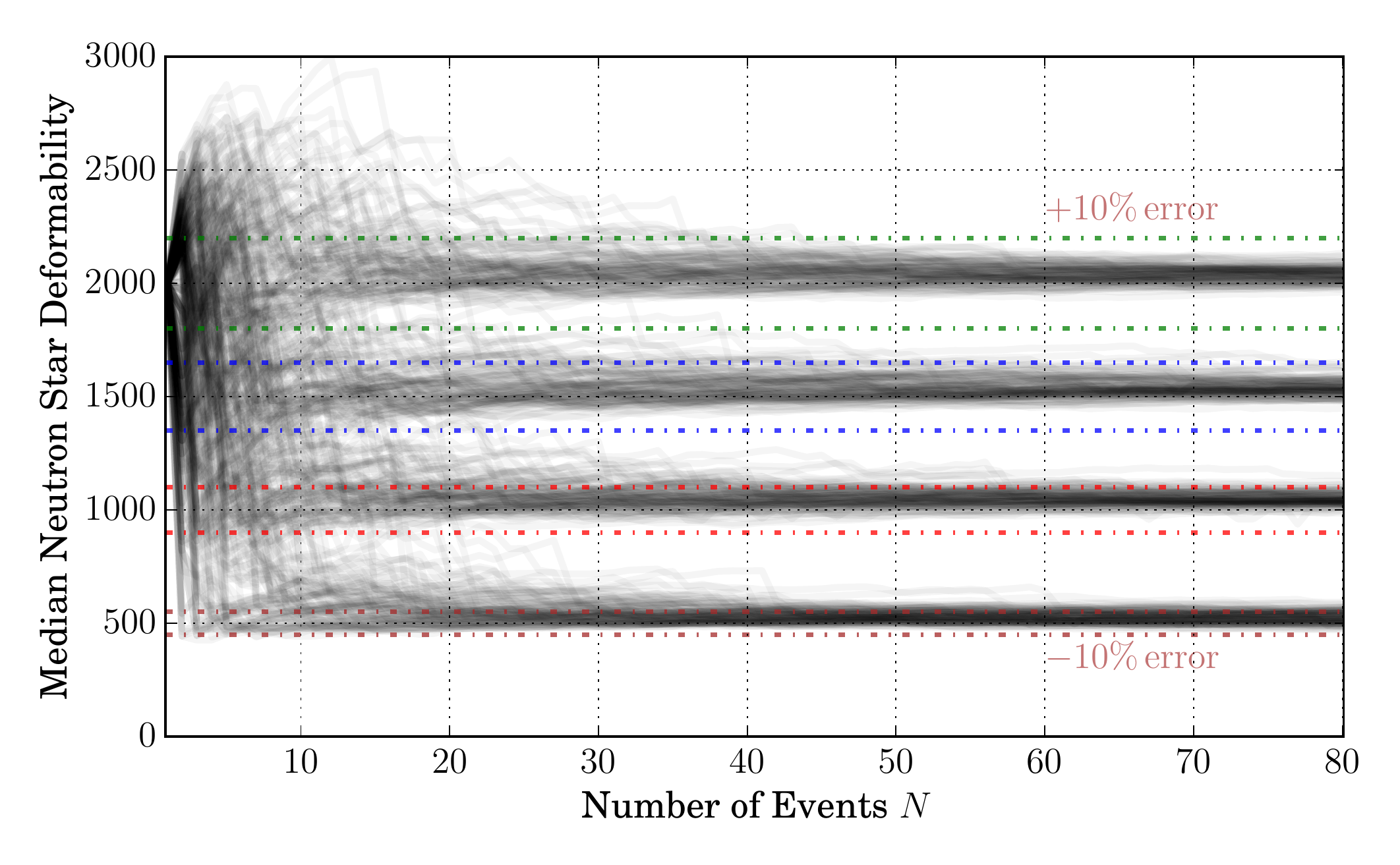}\\
\includegraphics[trim=1cm 0 0 0, width=1.025\columnwidth]{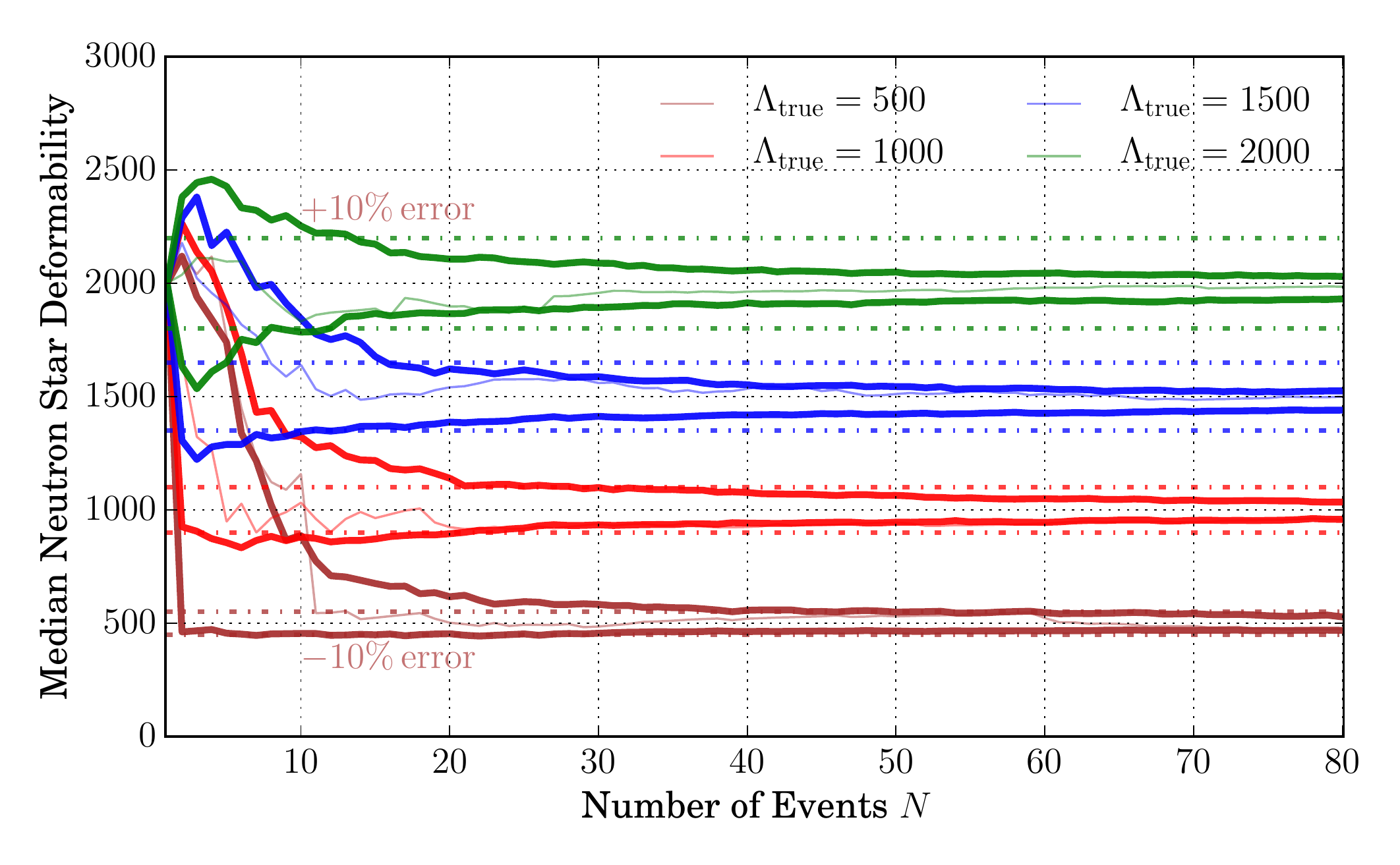}
\includegraphics[trim=0 0 1cm 0, width=1.025\columnwidth]{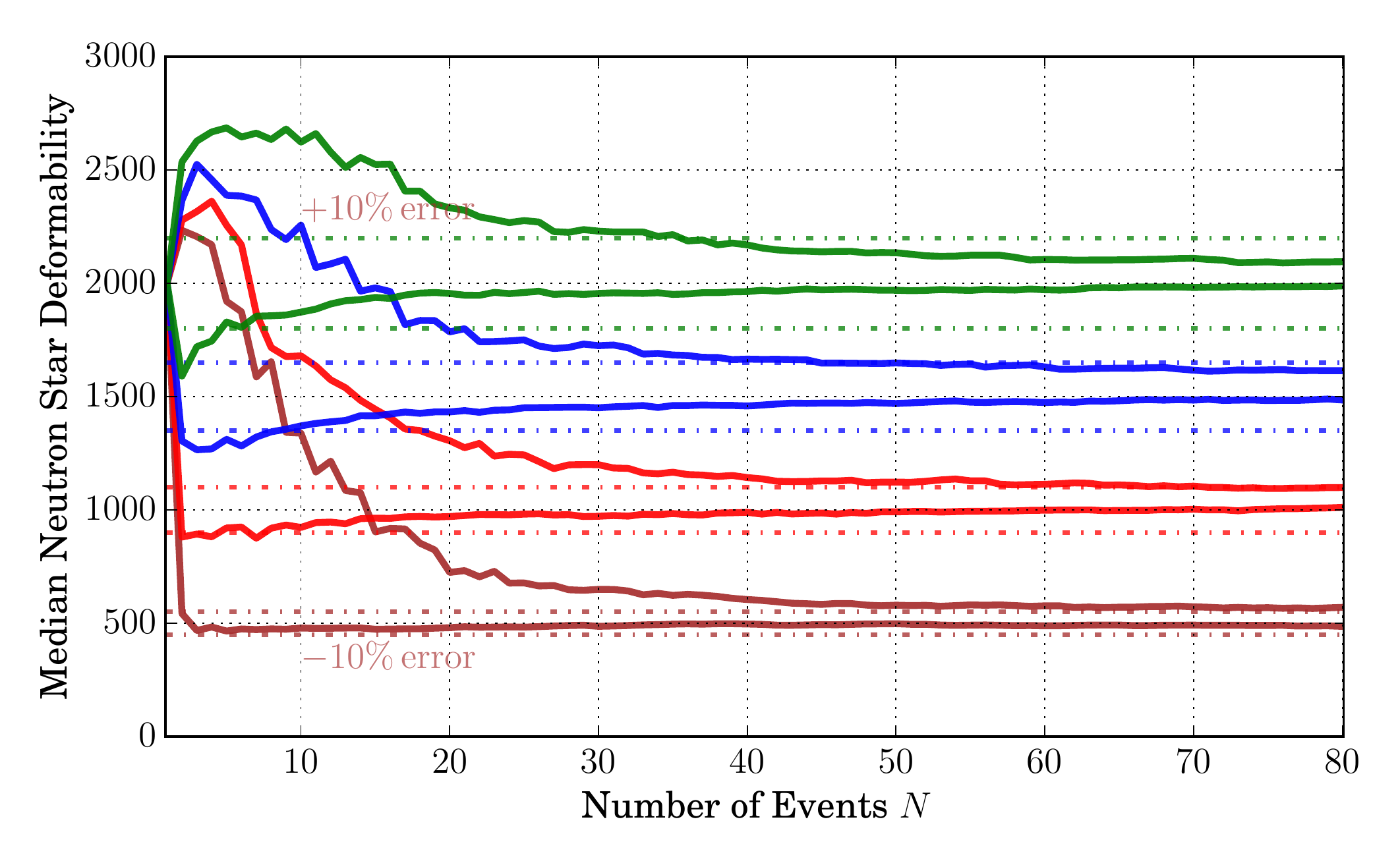}\\
\caption{
{\bf No Mass-Gap}, {\it top left}: The top figure shows the median value of the recovered
probability distribution for $\lambdans$, as a function of the number of observed 
events $N$. There are four ensembles of curves,
corresponding to $\lambdans=\{500,1000,1500,2000\}$, with a hundred
independent population draws within each ensemble. One curve in each ensemble
is highlighted in color, representing the realizations already plotted in
Fig.~\ref{fig:TT_Lambda_vs_N_CI90_0}.
In the same color we show $\pm 10\%$ error-bounds on $\lambdans$ with
horizontal dash-dotted lines.
{\bf No Mass-Gap}, {\it bottom left}: Here we show the interval of $\lambdans$ values within
which the median $\lambdans$ lies for $90\%$ of the populations in
each ensemble shown in the top left panel.
We observe that within $10-25$ observations, the median of the measured 
cumulative probability distribution for $\lambdans$ converges to within $10\%$
of its true value.
{\bf Mass-Gap}, {\it right column}: These panels are identical to their counterparts on the left,
with the only difference that the BH masses in each population are restricted
to lie {\it outside} the astrophysical mass-gap (i.e. paradigm B). The
difference that
we observe under this paradigm is that we need more ($30+$) events to achieve 
the same ($10\%$) measurement accuracy for populations with $\lambdans<1000$.
For more deformable neutron stars, $10-25$ events would suffice.
}
\label{fig:TT_LambdaMedian_vs_N_AllInOne}
\end{figure*} 
\begin{figure*}
\centering    
% \textbf{No Mass-Gap \hspace{6cm} Mass-Gap}\par\medskip
\includegraphics[width=1.025\columnwidth,trim=1cm 0 0 0]{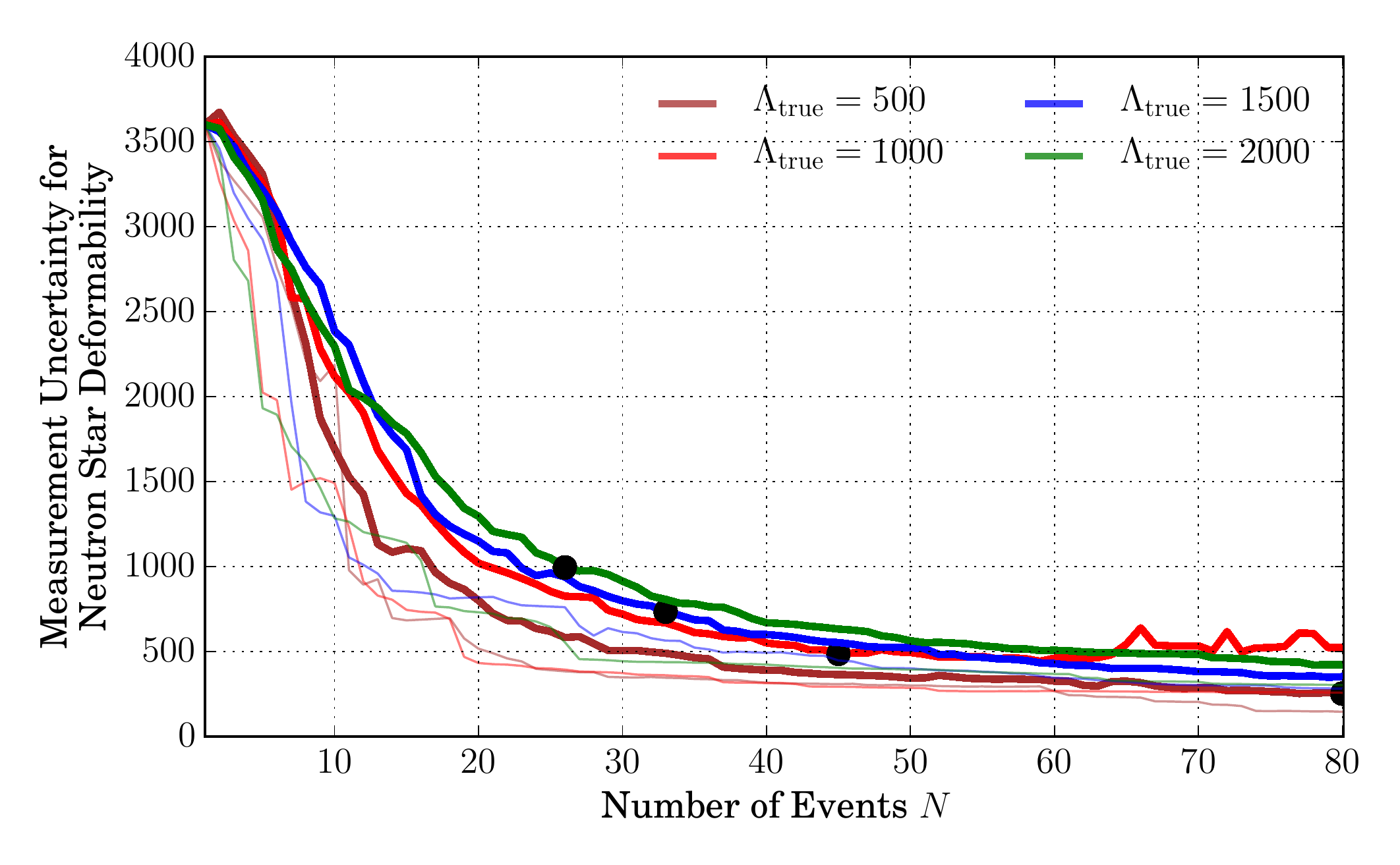}
\includegraphics[width=1.025\columnwidth,trim=0 0 1cm 0]{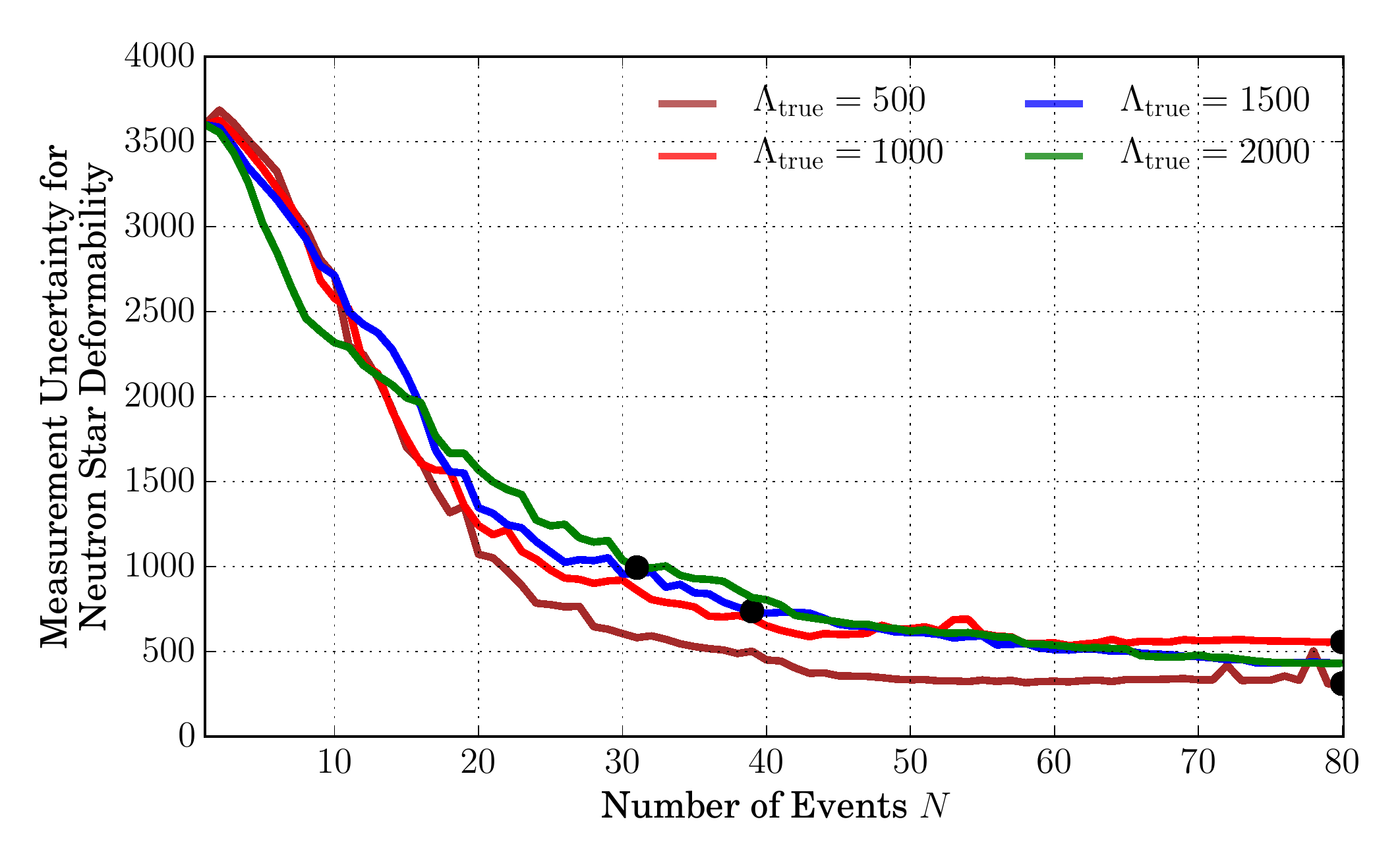}
\caption{{\bf No Mass-Gap} ({\it left}): This panel shows the width of $\lambdans$ interval within
which the $90\%$ credible intervals for $\lambdans$ lie, for $90\%$ of 
the populations in each ensemble, as a function of the number of observed events
$N$. Details of how this is calculated are given in the text.
The populations are sampled under paradigm A, which allows BH masses to
fall within the astrophysical mass-gap.
Each panel corresponds to a unique value of populations' $\lambdans$,
decreasing from $2000\rightarrow 500$ as we go from top to bottom.
One curve in each ensemble is highlighted in color (thin lines), representing the 
realizations already plotted in Fig.~\ref{fig:TT_Lambda_vs_N_CI90_0}.
{\bf Mass-Gap} ({\it right}): This panel shows populations drawn under paradigm B, which
respects the mass-gap.
We find that with approximately $25$ or so events, we begin to put
statistically meaningful constraints on $\lambdans$, restricting it to within
$\pm 50\%$ of the true value. We can expect to achieve this with a few years
of design aLIGO operation~\cite{Abadie:2010cfa}. Further tightening of 
$\lambdans$ credible intervals will require $40+$ events.
}
\label{fig:TT_LambdaError_vs_N_L500_2000_CI90_0_AllInOne}
\end{figure*}
\begin{figure*}
\centering    
% \textbf{No Mass-Gap \hspace{6cm} Mass-Gap}\par\medskip
\includegraphics[width=1.025\columnwidth,trim=1cm 0 0 0]{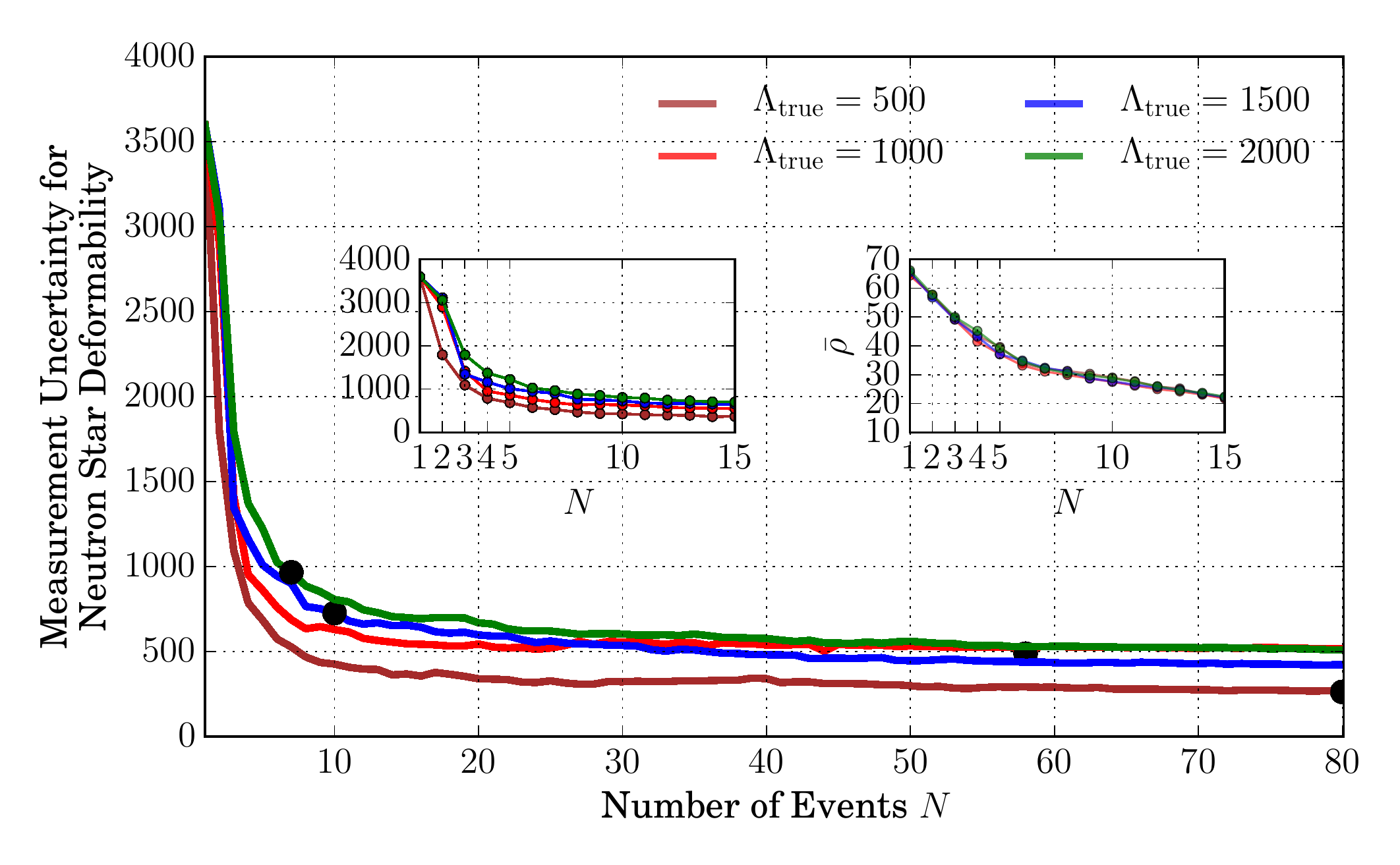}
\includegraphics[width=1.025\columnwidth,trim=0 0 1cm 0]{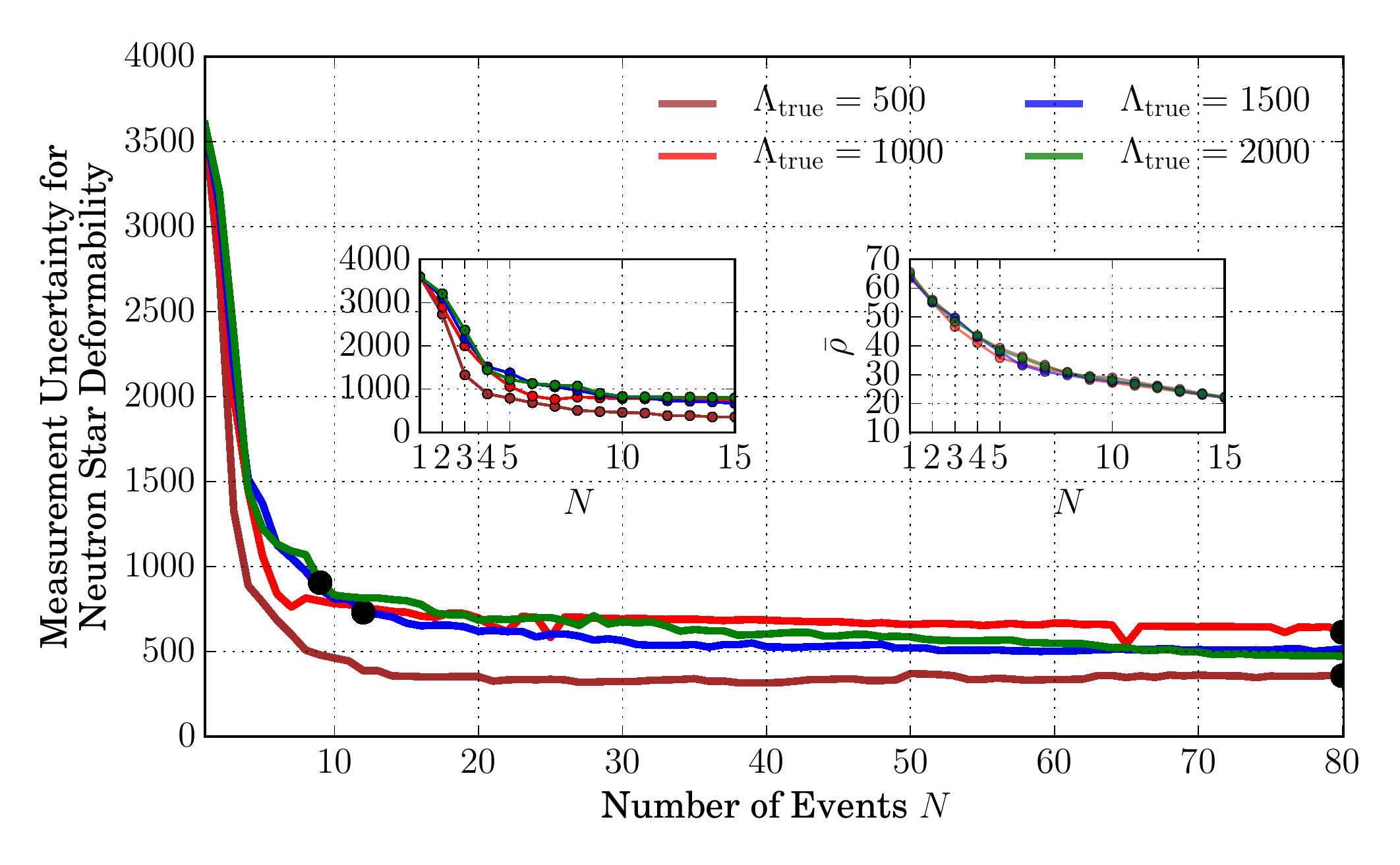}
\caption{This figure is similar to Fig.~\ref{fig:TT_LambdaError_vs_N_L500_2000_CI90_0_AllInOne},
with the only difference being that events in each population
have been sorted according to their signal strength (SNR), instead of
their simulated chronology. We note that information about
the tidal deformability of neutron stars comes primarily 
from the loudest $5-10$ events, whether we allow BH masses
in the mass gap (left panel) or restrict them to
$\mbh\geq 5M_\odot$ (right panel). Left inset zooms
in on the main figure for the first $15$ events. Right inset
shows the actual (ensemble mean) SNR value for each event. We find that
events with $\rho\gtrsim 20-30$ provide the bulk of tidal information
in our analysis.
}
\label{fig:TT_LambdaError_vs_N_L500_2000_CI90_0_AllInOne_SNRSorted}
\end{figure*}

In the previous section, we showed that single observations of NSBH
coalescences at moderate SNRs have little information about the internal
structure of neutron stars that will be accessible to Advanced LIGO at its
design sensitivity. We expect all neutron stars to share the same equation of
state, and hence the same $\lambdans(\mns)$. In addition, we know that the mass
distribution of (most) NSs that have not been spun up to millisecond periods
(which are the ones we focus on in this paper, by setting $\chins\approx 0$) is
narrowly peaked around $\sim 1.35M_\odot$~\cite{Kiziltan2013}. Therefore,
information from multiple NSBH observations can be combined to improve our
estimation of $\lambdans$. We explore the same in this section within a fully
Bayesian framework. We refer the reader to Ref.~\cite{Mandel:2009pc,Lackey2014,
Wade:2014vqa} for similar analyses of BNS inspirals.

% \textbf{Multiple identical sources at low SNR: }\label{s2:identical_multiple}
% 
An intuitive understanding of the problem is gained by considering first
multiple {\it identical} sources with realistic but different SNRs. Let us consider the case
of a population of optimally oriented binaries~\footnote{An optimally oriented
binary is one which is located directly overhead the detector, with the 
orbital angular momentum parallel to the line joining the detector to the
source. Such a configuration maximizes the observed GW signal strength in 
the detector.}, distributed uniformly in spatial volume out
to a maximum {\it effective} distance~\footnote{{\it effective} distance $D$ 
is a combination of distance to the source, its orientation, and its sky
location angles; and has a one-to-one correspondence with SNR for non-precessing
sources. This is so because for such sources, their location and orientation
remain constant over the timescales within which they sweep through
aLIGO's sensitive frequency band.}.
$D^\mathrm{max}$. $D^\mathrm{max}$ is set by the minimum SNR 
threshold $\rho_\mathrm{min}$ at which a source is considered
detectable~\footnote{which
we take as $\rho_\mathrm{min}=10$ throughout.}. Next, we divide this volume into $I$
concentric shells, with radii $D_i$. If we have a measurement error
$\sigma_0$ for $\lambdans$, associated with a source located at $D=D_0$,
the same error for the same source located within the $i-$th shell would
be $\sigma_i=\sigma_0 \dfrac{D_i}{D_0}$. Ref.~\cite{Markakis:2010mp}
calculated that the combined error $\sigma$ from $N$ independent
measurements of $\lambdans$ in such a setting to be
\begin{align}\label{eq:1oversigma}
\frac{1}{\sigma^2} =& \sum_{i=1}^I \frac{N_i}{\sigma_i^2} = \left(\frac{D_0}{\sigma_0}\right)^2 \sum_{i=1}^I\frac{N_i}{D_i^2}\\ \nonumber =& \left(\frac{D_0}{\sigma_0}\right)^2 \int_0^{D^\mathrm{max}} \dfrac{4\pi D^2 n}{D^2}\D D = \left(\frac{D_0}{\sigma_0}\right)^2 \dfrac{3N}{(D^\mathrm{max})^2},
\end{align}
where $N_i$ is the number of sources within the $i-$th shell (s.t.
$N:=\sum N_i$), and $n$ is the number density of sources in volume.
The root-mean-square (RMS) averaged measurement error from $N$ sources is 
then~\cite{Markakis:2010mp}
\begin{equation}\label{eq:rmsSigmaIdenticalSources}
 \sigma_{avg} := \frac{1}{\sqrt{1/\sigma^{2}}} = \frac{\sigma_0}{D_0} D^\mathrm{max} \frac{1}{\sqrt{3 N}},
\end{equation}
given a fiducial pair $(\sigma_0, D_0)$. It is straightforward to deduce from
Eq.~\ref{eq:rmsSigmaIdenticalSources} that measurement uncertainty scales as 
$1/\sqrt{N}$, and the uncertainty afforded by a single observation with a high
SNR $\rho_c$ can be attained with $N = \rho_c^2/300$ realistic observations
that have $\rho\geq\rho_\mathrm{min}$. E.g., to get to the
level of certainty afforded by a single observation with $\rho=70$, we would
need $49/3\approx 16-17$ realistic (low SNR) detections.

While we discussed Eq.~\ref{eq:rmsSigmaIdenticalSources} for a population
of optimally oriented sources, it is valid for a more general population
distributed uniformly in effective volume~\cite{Markakis:2010mp}
($\propto D^3$).
However, it 
still only applies to sources with identical masses and spins, and we 
overcome this limitation by performing a fully Bayesian analysis next.

\textbf{Astrophysical source population: }\label{s2:astro_multiple}
Imagine that we have $N$ stretches of data, $d_1, d_2, \cdots, d_N$, each 
containing a single signal emitted by an NSBH binary. Each of these signals can
be characterized by the non-tidal source parameters
$\vec{\theta} := \{\mbh, \mns, \chibh, \chins, \vec{\alpha}\}$,
and $\{\lambdans\}$, where $\vec{\alpha}$ contains extrinsic parameters,
such as source distance, inclination, and sky location angles.
As before, let $H$ denote all of our collective prior knowledge; for instance,
$H$ includes our assumption that all NSs in a single population have the same
deformability parameter $\lambdans$, and that its cumulative measurement is
therefore possible.
The probability distribution for $\lambdans$, given $N$ unique and
independent events, is
\begin{eqnarray}
 p(\lambdans |&& \hspace{-4mm}d_1, d_2, \cdots, d_N, H)\hspace{50mm}\nonumber\\ &=& \dfrac{p(d_1,d_2,\cdots,d_N |\lambdans , H)\,p(\lambdans|H)}{\int p(\lambdans |H) p(d_1,d_2,\cdots,d_N |\lambdans , H)\D\lambdans},\label{eq:p11}\\
  &=& \dfrac{p(\lambdans|H) \prod_{i} p(d_i|\lambdans, H)}{\int p(\lambdans ) p(d_1,d_2,\cdots,d_N |\lambdans , H)\D\lambdans},\label{eq:p12} \\
  &=& \dfrac{p(\lambdans|H) \prod_{i} \left( p(\lambdans |d_i, H)\dfrac{p(d_i)}{p(\lambdans|H)} \right)}{\int\, p(\lambdans|H )\, p(d_1,d_2,\cdots,d_N |\lambdans , H)\D\lambdans}\label{eq:p13};
\end{eqnarray}
where Eq.~\ref{eq:p11} and Eq.~\ref{eq:p13} are application of Bayes' theorem,
while Eq.~\ref{eq:p12} comes from the mutual independence of all events.
Assuming in addition that all events are {\it equally likely}: 
$p(d_i) = p(d_j) = p(d)$, we get
\begin{eqnarray}
 p(&&\hspace{-4mm}\lambdans | d_1, d_2, \cdots, d_N, H)\hspace{50mm}\nonumber\\
%   &=& \dfrac{p(\lambdans) \left(\dfrac{p(d)}{p(\lambdans)}\right)^N\prod_{i} p(\lambdans |d_i, H) }{\int p(\lambdans ) p(d_1,d_2,\cdots,d_N |\lambdans , H)\D\lambdans}\label{eq:p21},\\
  &=& p(\lambdans)^{1-N}\times \dfrac{p(d)^N}{\int p(\lambdans) p(d_1, d_2, \cdots, d_N |\lambdans, H)\D\lambdans}\nonumber\\ &&\hspace{3mm}\times\prod_i p(\lambdans |d_i, H)\label{eq:p22},
\end{eqnarray}
where the prior probability $p(\lambdans|H)$ is written $p(\lambdans)$ for
brevity. {\it A priori}, we assume that no particular value of $\lambdans$ is
preferred over another within the range $[0, 4000]$, i.e.
\begin{equation}\label{eq:lprior}
 p(\lambdans | H) = \dfrac{1}{4000}\,\mathrm{Rect}\left(\frac{\lambdans-2000}{4000}\right).
\end{equation}
With a uniform prior, the first two factors in Eq.~\ref{eq:p22} can be
absorbed into a normalization factor $\mathcal{N}$, simplifying it to
\begin{equation}\label{eq:lambdaMultiple}
 p(\lambdans | d_1, d_2, \cdots, d_N; H) = \mathcal{N}\prod_{i=1}^N p(\lambdans | d_i, H).
\end{equation}
In the second set of terms in Eq.~\ref{eq:lambdaMultiple} (of the form 
$p(\lambdans | d_i, H)$), each is the probability distribution for $\lambdans$
inferred {\it a posteriori} from the \textit{i}-th observation by marginalizing
\begin{equation}\label{eq:margpost}
 p(\lambdans | d_i, H) = \int\, p(\vec{\theta}, \lambdans | d_i, H)\, \D \vec{\theta},
\end{equation}
where $p(\vec{\theta}, \lambdans | d_i, H)$ is the inferred joint probability 
distribution of all source parameters $\vec{\theta}\cup\{\lambdans\}$ for the 
$i$-th event, as given by Eq.~\ref{eq:postprob}. We note that 
Fig.~\ref{fig:SingleSystemLambdaPDFvsSNR} illustrates $p(\lambdans | d_i, H)$
for three individual events. By substituting
Eq.~\ref{eq:lprior}-\ref{eq:margpost} into Eq.~\ref{eq:lambdaMultiple}, we
calculate the probability distribution for $\lambdans$ as measured using $N$
independent events.

Our goal is to understand the improvement in our measurement of $\lambdans$
with the number of recorded events. To do so, we simulate a population~\footnote{%
A population here is an ordered set of events, and an event itself is the 
set of parameters describing one astrophysical NSBH binary.}
of $N$ events, and quantify what we learn from each successive observation 
using Eq.~\ref{eq:lambdaMultiple}. This allows us to quantify how
rapidly our median estimate for $\lambdans$ converges to the true value,
and how rapidly our credible intervals for the same shrink, with increasing
$N$. Finally, we generate and analyze an ensemble of populations in order to
average over the stochastic process of population generation itself.

In order to generate each population, the first step is to fix
the NS properties: (i) NS mass $\mns=1.35M_\odot$, (ii) NS spin $\chins=0$
and (iii) NS tidal deformability $\lambdans=$ fixed value chosen from
$\{500,800,1000,1500,2000\}$. Next, we generate events, by sampling BH mass
(uniformly) from $\mbh\in[3M_\odot,6.75M_\odot]$, BH spin (uniformly)
from $\chibh\in[0, 1]$, orbital inclination from $\iota\in[0,\pi]$, and 
source location uniform in spatial volume\footnote{with a minimum SNR 
$\rho_\mathrm{min}=10$}.
We restrict ourselves to positive aligned BH spins, since binaries with
anti-aligned spins have very little information to add at realistic SNRs,
as demonstrated in Fig.~\ref{fig:TT_LambdaCIWidths90_0_Lambda_SNR}. This is
to be taken into account when the number of observations is related to detector
operation time. 
We repeat this process till we have an ordered set of $N$ events.
Since we want to analyze not just a single realization of an astrophysical
population, but an ensemble of them, we make an additional approximation to
mitigate computational cost. Complete Bayesian parameter estimation is
performed for a set of simulated signals whose parameters are the vertices
of a regular hyper-cubic grid (henceforth ``G'') in the space of
$\{q\}\times\{\chibh\}\times\{\rho\}$, with each sampled at $q=\{2,3,4,5\}$,
$\chibh=\{-0.5,0,0.5,0.75\}$, and $\rho=\{10,20,30,50,70\}$.
All events in each population draw are substituted by their respective nearest
neighbours on the grid G.
Our chosen signal parameter distribution is different from some other studies
in literature, which often sample from more astrophysically motivated population
distribution functions~\cite{Mandel:2009pc}. We chose one that is sufficiently
agnostic in absence of actual known NSBHs, and pragmatic enough for generating
population ensembles.

In Fig.~\ref{fig:TT_Lambda_vs_N_L800_CI90_0} we show illustrative results
for a single population with neutron star deformability $\lambdans=800$.
In the top panel, each curve shows the 
probability distributions for $\lambdans$ as inferred from $N$ events, with $N$
ranging from $1-80$. We also mark the $90\%$ credible intervals associated
with each of the probability distribution curves. The first few observations
do not have enough information to bound $\lambdans$ much more than
our prior from Eq.~\ref{eq:lprior} does. 
In the bottom panel, we present information derived from the top panel.
The line-circle curve shows the measured median value from $N$ observations.
The pair of dashed (dotted) horizontal lines mark
$\pm25\%$ ($\pm50\%$) error bars. At each $N$, the range spanned by the 
filled region is the $90\%$ credible interval deduced from the same 
events. This figure somewhat quantifies the qualitative deductions we made
from the left panel. We find that the median does track the true value quickly,
reaching within its $10\%$ with $10-15$ observations. This is as one expects of 
injections in zero noise where random fluctuations are unable to shift the
median away from the true value, so long as the measurement is not restricted
by the prior. With the same information, our credible intervals also shrink to $\pm 25\%$.
In Fig.~\ref{fig:TT_Lambda_vs_N_CI90_0} we show further results from four
independent populations for $\lambdans=\{500,1000,1500,2000\}$. As in the 
right panel of Fig.~\ref{fig:TT_Lambda_vs_N_L800_CI90_0}, the line-circle curves
track the median $\lambdans$, while the filled regions show
the associated $90\%$ credible intervals. From the figure, we observe
the following: (i) the shrinkage of credible interval widths with increasing
$N$ happens in a similar manner for each $\lambdans$, 
% (ii) within $\sim 10$ observations, median $\lambdans$ for all populations lie
% within $10\%$ of the true values, 
and (ii) it takes
approximately $20$ events to distinguish definitively (with $90\%$ credibility)
between deformable NSs with $\lambdans=2000$ and compact NSs with 
$\lambdans=500$, or equivalently to distinguish between hard, moderate and soft
nuclear equations of state. This is comparable to what has been found for
binary neutron stars~\cite{DelPozzo:13,Chatziioannou:2015uea,Agathos:2015a}.

So far we have discussed individual realizations of NSBH populations. The 
underlying stochasticity of the population generation process makes it
difficult to draw generalized inferences (from a single realization of an
NSBH population) about the measurability of $\lambdans$. In order to mitigate
this, we discuss ensembles of population draws next. In
Fig.~\ref{fig:TT_LambdaMedian_vs_N_AllInOne} we show the median
$\lambdans$ as a function of the number of observed
events, for four population ensembles, with a hundred population draws
in each ensemble. Lets focus on the {\it top left} panel first. In it, we
show the median $\lambdans$ for all populations in four ensembles,
with true $\lambdans=\{2000,1500,1000,500\}$ from top to bottom.
Populations highlighted in color are simply those that we discussed in
Fig.~\ref{fig:TT_Lambda_vs_N_CI90_0}. Dash-dotted horizontal lines
demarcate $\pm10\%$ error intervals around the true $\lambdans$ values.
The panel just below it shows the range of $\lambdans$ that encloses the
median $\lambdans$ for $90\%$ of the populations in {\it each}
ensemble. In other words, this panel shows the range of $\lambdans$ within which
the median $\lambdans$ value for $90\%$ of NSBH populations is
expected to lie. From these panels, we observe that our median $\lambdans$
values will be within $10\%$ of the {\it true} value after $\sim 25$
detections of less deformable neutron stars ($\lambdans\leq 1000$), or
after as few as $15$ detections of more deformable neutron stars
($\lambdans\geq 1500$). This is not surprising because we inject simulated
signals in {\it zero} noise, which ensures that the median not be shifted away
from the true value. That it takes $15+$ events for the median to approach
the true value is a manifestation of the fact that the measurement is limited
by the prior on $\lambdans$ when we have fewer than $15$ events.
The results discussed in Fig.~\ref{fig:TT_Lambda_vs_N_L800_CI90_0},
\ref{fig:TT_Lambda_vs_N_CI90_0} and the left two panels
of Fig.~\ref{fig:TT_LambdaMedian_vs_N_AllInOne} apply to the parameter
distribution spanned by the grid G. This distribution allows for $\mbh$
as low as $2.7M_\odot$ (i.e. $q=2$).
Given that disruptive signatures are strongest for small $\mbh$, we now
investigate an alternate paradigm in which no black hole masses fall within
the mass gap $2-5M_\odot$ suggested by astronomical
observations~\cite{Bailyn:1997xt,Kalogera:1996ci,Kreidberg:2012,
Littenberg:2015tpa}. We will henceforth denote our standard paradigm, which
does not respect the mass-gap, as paradigm A; with paradigm B being
this alternate scenario.
Both right panels of the figure are identical to their corresponding left
panels, but drawn under population paradigm B. Under this paradigm, we
expectedly find that information accumulation is much slower. It would
take $25-40$ detections with $\rho\geq10$ under this paradigm, for our median
$\lambdans$ to converge within $10\%$ of its true value.
% % 

Finally, we investigate the statistical uncertainties associated with
$\lambdans$ measurements. We use $90\%$ credible intervals as our measure of
the same. First, we draw an ensemble of a hundred populations each for
$\lambdans=\{500,1000,1500,2000\}$. For each population $i$ in each ensemble,
we construct
its $90\%$ credible interval $[{\lambdans^{90\%}}_{i-},{\lambdans^{90\%}}_{i+}]$.
Next, we construct the interval $[X^-,Y^-]$ that contains ${\lambdans^{90\%}}_{i-}$
for $90\%$ of the populations in each ensemble; and similarly $[X^+,Y^+]$
for ${\lambdans^{90\%}}_{i+}$. Finally, in the left panel of 
Fig.~\ref{fig:TT_LambdaError_vs_N_L500_2000_CI90_0_AllInOne}, we show the
conservative width $|Y^+ - X^-|$ that contains the $90\%$ credible
intervals for $90\%$ of all populations in each ensemble~\footnote{Drawn
under paradigm A (mass-gap not respected).}.
From top to bottom, the population $\lambdans$
decreases from $\lambdans=2000\rightarrow 500$, corresponding to decreasingly
deformable NSs with softer equations of state. We observe the following:
(i) for moderately-hard to hard equations of state with $\lambdans\geq 1000$,
we can constrain $\lambdans$ within $\pm 50\%$ using only $10-20$ events, and
within $\pm 25\%$ (marked by black circles) with $25-40$ events; (ii) for softer 
equations of state with
$\lambdans<1000$, we will achieve the same accuracy with $20-30$ and $50+$ 
events, respectively; and (iii) for the first $5$ or so observations, our
measurement spans the entire prior allowed range:
$\lambdans\in[0,4000]$, as shown by the plateauing of the $90\%$ 
credible intervals towards the left edge to $90\%$ of $4000$, i.e. $3600$.
The right panel in Fig.~\ref{fig:TT_LambdaError_vs_N_L500_2000_CI90_0_AllInOne}
is identical to the left one, with the difference that populations are drawn
under paradigm B, which does {\it not} allow for BH masses to fall within
the mass-gap. We find that
for NSs with $\lambdans\leq 1000$, it would take $25-40$ events to
constrain $\lambdans$ within $\pm 50\%$ and $50+$ events to constrain it
within $\pm 25\%$. This is somewhat slower than paradigm A, as is to be
expected since here we preclude the lowest mass-ratios, which correspond to
signals with largest tidal signatures. For $\lambdans>1000$ we find that we
can constrain $\lambdans$ within $\pm 50\%$ with a similar number of events as
for paradigm A, but will need more ($30-40$, as compared to $25-40$) events
to further constrain it to within $\pm 25\%$ of the true value.
Under either paradigms, we find that measuring $\lambdans$ better than
$25\%$ will require $\mathcal{O}(10^2)$ observations of disruptive NSBH
mergers.
These results demonstrate that our probed set of disruptive NSBH mergers is as
informative of NS tidal properties as are BNS populations, if we assume a uniform mass
distribution for NSs and zero NS spins, and possibly more if NS masses are
not distributed uniformly~\cite{Agathos:2015a}.
However, being a subset of all NSBH binaries, the accumulation of information
from NSBH signals in general may be slower than from binary neutron stars,
depending on the distribution of BH masses in coalescing NSBH binaries.

{\it Do all events matter?} In Ref.~\cite{Lackey2014} the authors demonstrate
that the overwhelming majority of information about the NS equation of state
comes from the loudest $\sim 5$ events in the case of binary neutron star
detections, and not from the combined effect of a large number of low-SNR
systems.
The question naturally arises if the same is true for NSBH sources as well.
Therefore, in Fig.~\ref{fig:TT_LambdaError_vs_N_L500_2000_CI90_0_AllInOne_SNRSorted}
we re-evaluate the accumulation of information with each successive NSBH 
detection, sorting the events in each population by their SNR instead of 
simulated chronology. We find the same qualitative behavior as in the case of 
BNSs~\cite{Lackey2014}. 
Whether or not there is an astrophysical mass gap, we find
that the bulk of tidal information will be furnished by the loudest $5-10$
NSBH detections of aLIGO detectors. {\it This result is especially encouraging
to NR follow-up efforts for GW detections, as we now know that only a handful
of loudest NSBH events (with SNRs $\rho\gtrsim 20-30$) are the ones that may merit
full numerical-GR + magnetohydrodynamical follow-up simulations.}

To summarize, in this section we study the improvement in our measurement of 
NS deformability parameter $\lambdans$ with an increasing number of events. We
do so by simulating plausible populations of disrupting NSBH binaries (with
$\rho\geq 10$). We find that:
(i) for more deformable neutron stars (harder equation of states), the median
value of $\lambdans$ comes within $10\%$ of the true value with as 
few as $10$ events, while achieving the same accuracy for softer equations of 
state will take $15-20$ source detections; (ii) the statistical uncertainty
associated with $\lambdans$ measurement shrinks to within $\pm50\%$ with
$10-20$ events, and to within $\pm 25\%$ with $50+$ events, when source 
$\lambdans\geq 1000$; (iii) for softer equations of state, the same could take
$25-40$ and $50+$ events, respectively for the two uncertainty thresholds;
and (iv) if BHs really do observe the astrophysical mass-gap, the information
accumulation is somewhat slower than if they do not. We conclude that within
$20-30$ observations, aLIGO would begin to place very interesting bounds on 
the NS deformability, which would allow us to rule out or rank different
equations of state for neutron star matter. Within this population, we also
find that it will be the loudest $5-10$ events that will furnish most of the
tidal information. Our key findings are summarized in 
Fig.~\ref{fig:TT_LambdaMedian_vs_N_AllInOne} -
\ref{fig:TT_LambdaError_vs_N_L500_2000_CI90_0_AllInOne_SNRSorted}.

%%%%%%%%%%%%%%%%%%%%%%%%%%%%%%%%%%%%%%%%%%%%%%%%%%%%%%%%%%%%%%%%%%%%%%%%%%%%%%%
\section{Discussion}\label{s1:discussion}
%%%%%%%%%%%%%%%%%%%%%%%%%%%%%%%%%%%%%%%%%%%%%%%%%%%%%%%%%%%%%%%%%%%%%%%%%%%%%%%

The pioneering observation of gravitational waves by Advanced LIGO
harbingers the dawn of an era of gravitational-wave astronomy where observations
would finally drive scientific discovery~\cite{Abbott:2016blz}. As confirmed by
the first observations~\cite{Abbott:2016blz,Abbott:2016nmj,Abbott:2016nhf},
stellar-mass compact binary mergers emit GWs right in the sensitive frequency
band of the LIGO observatories, and are their primary targets.
Neutron star black hole binaries form a physically distinct sub-class of
compact binaries. We expect to detect the first of them in the upcoming
observing runs~\cite{Abbott:2016ymx}, and subsequently at a healthy rate of
$0.2-300$ mergers a year when aLIGO detectors reach design
sensitivity~\cite{Abadie:2010cf}.

NSBH binaries are interesting for various reasons. Unlike BBHs, the presence of
matter allows for richer phenomena to occur alongside the strong-field
gravitational dynamics. The quadrupolar moment of the NS changes during the
course of inspiral, which increases the inspiral rate of the binary and alters the
form of the emitted gravitational waves. Close to merger, under restricted but
plausible conditions, the neutron star is disrupted by the tidal field of its 
companion black hole and forms an accretion disk around it. This disruption
reduces the quadrupolar moment of the system, and decreases the amplitude of
the emitted GWs from the time of disruption through to the end of ringdown.
Both of these phenomena are discernible in their gravitational-wave signatures
alone. In addition, if the neutron star matter is magnetized, the magnetic
winding above the remnant black hole poles can build up magnetic fields
sufficiently to power short gamma-ray bursts (SGRB)~\cite{Foucart:2015a,
Lovelace:2013vma,Deaton2013,Foucart2012,Shibata:2005mz,Paschalidis2014}.
Therefore a coincident observation of gravitational waves from an NSBH merger
and a SGRB can potentially confirm the hypothesis that the former is a
progenitor of the latter~\cite{eichler:89,1992ApJ...395L..83N,moch:93,
Barthelmy:2005bx,2005Natur.437..845F,2005Natur.437..851G,Shibata:2005mz,
Tanvir:2013,Paschalidis2014}.

In this paper we study the observability of tidal signatures in the
gravitational-wave spectrum of NSBH binaries. More specifically, we investigate
three questions. First, what is the effect of not including tidal effects in 
templates while characterizing NSBH signals? Second, if we do include tidal 
effects, how well can we measure the tidal deformability of the NS
(parameterized by $\lambdans$) from individual NSBH signals? And third, as we
observe more and more signals, how does our knowledge of $\lambdans$ improve?
In the following, we summarize our main findings.

First, we study the effects of not including tidal terms in our search
templates while characterizing NSBH signals. We expect that the waveform
template that best fits the signal would compensate for the reduced number of
degrees of freedom in the template model by moving away from the true
parameters of the binary. This should result in a {\it systematic} bias in 
the recovered values of non-tidal source parameters, such as its masses 
and spins. In order to quantify it, we inject tidal signals into zero noise,
and perform a Bayesian parameter estimation analysis on them using templates
{\it without} tidal terms.
We use the LEA+ model (c.f. Sec.~\ref{s2:waveforms}) to produce tidal waveforms
that incorporate the effect
of NS distortion during inspiral, and of its disruption close to merger. Our
injected signals sample the region of NSBH parameter space where NS disruption
prior to binary merger is likely {\it and} can be modeled using LEA+. Their
parameters are given by combinations of $q=\mbh/\mns=\{2,3,4,5\},
\chibh=\{-0.5, 0, 0.5, 0.75\}$ and $\lambdans=\{500,800,1000,1500,2000\}$.
Other parameters, such as source location and orientation, that factor out of
$h(t)$ as amplitude scaling are co-sampled by varying $\rho=\{20,30,50,70\}$.

At low to moderate SNRs ($\rho\lesssim 30$), we find that using BBH templates
does not significantly hamper our estimation of non-tidal parameters for NSBH
signals. In the worst case, when the BH mass is within the astrophysical 
mass-gap~\cite{Bailyn:1997xt,Kalogera:1996ci,Kreidberg:2012,Littenberg:2015tpa}
and its spin is positive aligned, the systematic biases in $\eta$ and $\chibh$
measurements do become somewhat comparable to statistical errors (ratio
$\sim 0.5-0.8$) under very restrictive conditions~\footnote{requiring a
companion BH with mass $\mbh\lesssim 4.5M_\odot$ (i.e. in the astrophysical
mass-gap), and the
hardest NS EoS considered (with $\lambdans\simeq 2000$).}, but never exceed them.
At high SNRs ($\rho\gtrsim 50$), systematic biases in $\mchirp$ become larger
than the statistical uncertainties. For $\eta$ and $\chibh$ the difference
is more drastic with the systematics reaching up to $4\times$ the statistical
errors. We therefore conclude that $\rho\simeq 30-50$ is loud enough to
motivate the use of tidal templates for even the estimation of non-tidal
parameters from NSBH signals.
We also conclude that low-latency parameter estimation algorithms, designed to
classify GW signals into electromagnetically active (NSBH and NSNS) and
inactive (BBH) sources, can use BBH templates to trigger GRB 
alerts~\cite{2012A&A...541A.155A,Singer:2014qca,Singer:2015ema,Pankow:2015cra,
Abbott:2016wya,Abbott:2016gcq} for NSBH signals with low to moderate SNRs
($\rho\lesssim 30$).
This is so because the primary requirement of identifying NS-X binaries (X =
\{NS, BH\}) can be achieved just as easily with BBH templates, on the basis of
the smaller component's mass\footnote{The smaller component mass is unlikely
to be significantly biased by missing tidal effects in filter templates below
$\rho\simeq 30$, as we show above.}.
We also speculate that NSBH detection searches are unlikely to be
affected by the choice of ignoring tidal effects in matched-filtering
templates, if these effects are too subtle to manifest in parameter estimation
below $\rho\simeq 30$.

% 
% % On the known NS and BH properties
% The number of neutron stars observed using conventional astronomical methods has
% grown rapidly in the recent past, both in isolated and two-body
% systems~\cite{Demorest:2010bx,Lyne:2004cj,2013Sci...340..448A,atnfcatalog,
% mcgillmagnetarcatalog,stellarcollapsemass}. Their masses span the range $1.2-2M_\odot$
% with an average of $\sim 1.35M_\odot$~\cite{Lattimer:2004sa,stellarcollapsemass}.
% On the other hand, NS spins have been observed to have magnitudes below
% $|\vec{S}_\mathrm{NS}|/\mns^2 < 0.01$~\cite{Miller:2014aaa}.
% % 
% On the other hand, indirect observations of stellar-mass BHs place their
% masses between $5-35M_\odot$, with their spin angular momenta 
% $|\vec{\chi}_\mathrm{BH}|$ ranging from small to nearly extremal (Kerr) values
% (see, e.g., Refs.~\cite{McClintockEtAl:2006,Miller:2009cw,Gou:2014una} for 
% examples of nearly extremal estimates of BH spins, Refs.~\cite{McClintock:2013vwa,
% Reynolds:2013qqa} for recent reviews of astrophysical BH spin measurements,
% and Figure 5 of Ref.~\cite{Miller:2014aaa} for a comparison of NS and BH spins).

Second, we turn the question around to ask: can we measure the tidal effects if
our template models did account for them? Tidal effects in our waveform model
are parameterized using a single deformability parameter 
$\lambdans\propto (R/M)_\mathrm{NS}^5$. In order to quantify the 
measurability of $\lambdans$, we inject the same tidal signals as before, and
this time perform a Bayesian analysis on them using {\it tidal} templates. 
The results are detailed in Sec.~\ref{s1:PEwithNS}.
At low SNRs ($\rho\simeq 20$), we find that the best we can do is to constrain
$\lambdans$ within $\pm 75\%$ of its true value at $90\%$ credible level. This
too only if the BH is spinning sufficiently rapidly, with $\chibh\gtrsim +0.7$,
and the NS has $\lambdans\gtrsim 1000$. At moderate SNRs ($\rho\simeq 30$), we
can constrain $\lambdans$ a little better, i.e. within $\pm 50\%$ of its true
value. This level of accuracy, however, again requires that BH spin
$\chibh\gtrsim+0.7$ and $\lambdans\gtrsim 1000$. Binaries with smaller BH spins
and/or softer NS EoSs will furnish worse than $\pm 75\%-\pm 100\%$ errors for
$\lambdans$. This trend continues as we increase the SNR from $\rho=30-50$. It
is not before we reach an SNRs as high as $\rho\simeq 70$ that we can shrink
$\lambdans$ errors substantially with a single observation (i.e. within
$\pm 25\%$ of its true value).
In summary, we find that with a single but moderately loud NSBH signal,
Advanced LIGO can begin to put a factor of $1-2\times$ constraints on NS tidal
deformability parameter. These constraints can subsequently be used to assess
the likelihood of various candidate equations of state for nuclear matter, and
possibly to narrow the range they span.

Third, knowing that single observations can furnish only so much information
about the NS equation of state, we move on to investigate how well we do with
multiple signals. In order to quantify how $\lambdans$ measurement improves
with the number of observed events $N$, we generate populations of NSBH signals
and combine the information extracted from each event.
The population generation procedure is as follows. The neutron star mass is
held fixed at $1.35M_\odot$, its spin at $\chins=0$, and its tidal
deformability is fixed to each of $\lambdans=\{500,1000,1500,2000\}$. Black hole
mass is sampled uniformly from the range $[2,5]\times 1.35=[2.7, 6.75]M_\odot$,
and spin from $\chibh\in[0,1]$. As before, our parameter choice here is given
by the intersection set of the mass range that allows for neutron star disruption
and the range supported by LEA+~\cite{Foucart2012,Foucart:2013a,Lackey:2013axa}.
In order to keep the computational cost reasonable, we make an additional
approximation. For every population generated, we replace the parameters of each
event by their nearest neighbor on the uniform grid G, which has vertices
at: $q=\{2,3,4,5\}\times\chibh=\{-0.5,0,0.5,0.75\}\times\lambdans=\{500,800,
1000,1500,2000\}\times\rho=\{10,20,30,50,70\}$.
This way, we only have to run full Bayesian parameter estimation analysis on
this fixed set of signals. 
There are two sources of error that enter the deductions we make from
a single population generated in the manner described above. First, since the
injection parameters are pushed to their nearest neighbor on a grid, we
find discrete jumps in $\lambdans$ errors as a function of $N$. And second, an
individual population is one particular realization of a stochastic process and
could have excursions that may never be found in another population. To
account for both of these limitations, we generate an ensemble of populations,
and conservatively combine information from all of them\footnote{See 
Sec.~\ref{s1:multiple_observations} for further details.}.

We probe two astrophysical paradigms, one that allows for BH masses to lie
within the astrophysical mass-gap (paradigm A), and one that does not (paradigm
B).
{\it For paradigm A}, we find the following: (i) for the softer equations of
state that result in less deformable neutron stars, $15-20$ detections bring
the measured probability distribution for $\lambdans$ entirely within the prior,
which ensures that the median $\lambdans$ tracks the true value to within $10\%$.
(ii) For NSBH populations with more deformable NSs ($\lambdans> 1000$),
the same is achievable within as few as $10$ (or $15$ at most) realistic
observations. (iii) The statistical uncertainty associated with $\lambdans$
measurement can be restricted to be within $\pm50\%$ using $10-20$ observations
when $\lambdans> 1000$), and using $25-40$ observations for softer equations
of state. All of the above is possible within a few years of design
aLIGO operation~\cite{Abadie:2010cfa}, if astrophysical BHs are allowed
masses $< 5M_\odot$ (i.e. in the mass-gap). However, further
restricting $\lambdans$ will require $50+$ NSBH observations.
{\it For paradigm B}, we find the information accumulation to be somewhat slower.
While the quantitative inferences for populations with $\lambdans>1000$ are
not affected significantly, we find that $\lambdans< 1000$ populations require 
$10-20\%$ more events to attain the same measurement accuracy as under
paradigm A. In either case, the accumulation of information from the general
NSBH population will likely be slower than from BNS inspirals~\cite{Mandel:2009pc,
Lackey2014,Wade:2014vqa,Agathos:2015a}, depending on the mass distribution of 
stellar-mass black holes. Though, template models for the latter may be more
uncertain due to missing point-particle PN terms at orders comparable to
tidal terms~\cite{Lackey2014}.
We conclude that within as few as $20-30$ observations of disruptive NSBH
mergers, aLIGO will begin to place interesting bounds on NS deformability.
This, amongst other things, will allow us to rank different equations of 
state for neutron star matter from most to least likely, within a few years'
detector operation.
We also find that, within this population, the loudest $5-10$ events (with SNRs
$\rho\gtrsim 20-30$) will provide us with most of the tidal information,
and will therefore merit full NR follow-up.
Our methods and results are detailed in Sec.~\ref{s1:multiple_observations}.

Finally, we note that the underlying numerical simulations used to calibrate
the waveform model used here have not been verified against
independent codes so far.
It is therefore difficult to assess the combined modeling error of LEA+ and its
effect
on our results. Our results here are, therefore, limited by the limitations of
our waveform model, and presented with this caveat. However, we do expect the
combined effect of modeling errors to {\it not} affect our {\it qualitative}
conclusions, especially since the underlying point-particle component of LEA+
includes all high-order terms, unlike past BNS studies~\cite{Lackey2014,
Wade:2014vqa}
In future, we plan to further the results presented here by using more recent
tidal models~\cite{Pannarale:2015jka,Hinderer:2016eia}, that
may improve upon LEA+\footnote{One of them~\cite{Pannarale:2015jka} is only an
amplitude model though, which has to be augmented with a compatible phase model
first.}.

%%%%%%%%%%%%%%%%%%%%%%%%%%%%%%%%%%%%%%%%%%%%%%%%%%%%%%%%%%%%%%%%%%%%%%%%%%%%%%%
% Acknowledgments
%%%%%%%%%%%%%%%%%%%%%%%%%%%%%%%%%%%%%%%%%%%%%%%%%%%%%%%%%%%%%%%%%%%%%%%%%%%%%%%
\begin{acknowledgments}
  We thank Benjamin Lackey, Francesco Pannarale, Francois Foucart, and Duncan Brown
  for helpful discussions. We gratefully acknowledge support
  for this research at CITA from NSERC of Canada, the Ontario Early 
  Researcher Awards Program, the Canada Research
  Chairs Program, and the Canadian Institute for Advanced Research.
  Calculations were performed at the Vulcan
  supercomputer at the Albert Einstein Institute;
  H.P. and P.K. thank the Albert-Einstein Institute,
  Potsdam, for hospitality during part of the time where this research
  was completed. M.P. thanks CITA for hospitality where part of the work
  was carried out.
\end{acknowledgments}

%%%%%%%%%%%%%%%%%%%%%%%%%%%%%%%%%%%%%%%%%%%%%%%%%%%%%%%%%%%%%%%%%%%%%%%%%%%%%%%
%%%%%%%%%%%%%%%%%%%%%%%%%%%%%%%%%%%%%%%%%%%%%%%%%%%%%%%%%%%%%%%%%%%%
\begin{appendix}

\section{Statistical uncertainty in measuring non-tidal parameters}\label{as1:nontidalerrors}
% 
% #################
\begin{figure*}
\centering 
\includegraphics[trim = {2cm 0 0 0},width=2.\columnwidth]{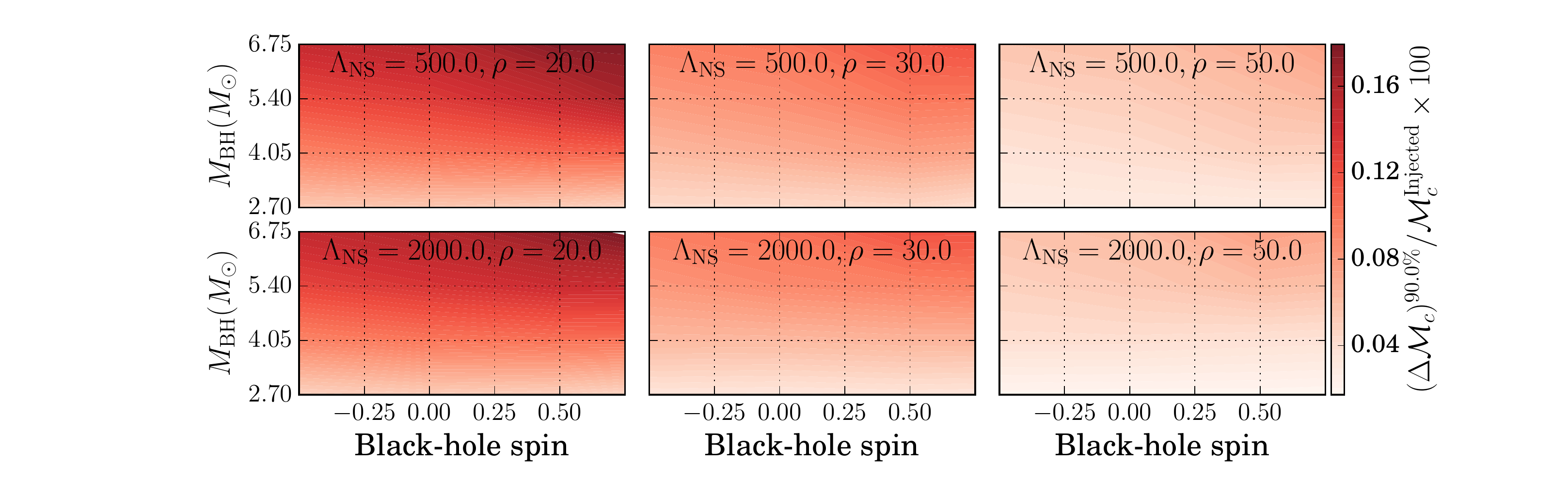}\\
\includegraphics[trim = {2cm 0 0 0},width=2.\columnwidth]{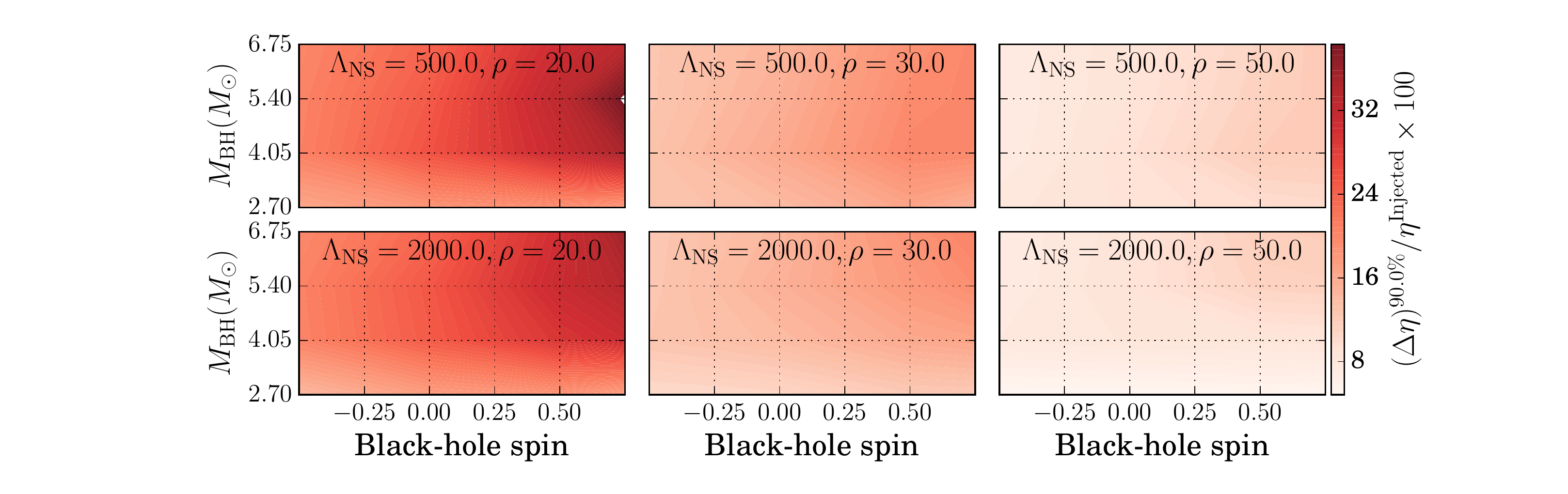}\\
\includegraphics[trim = {2cm 0 0 0},width=2.\columnwidth]{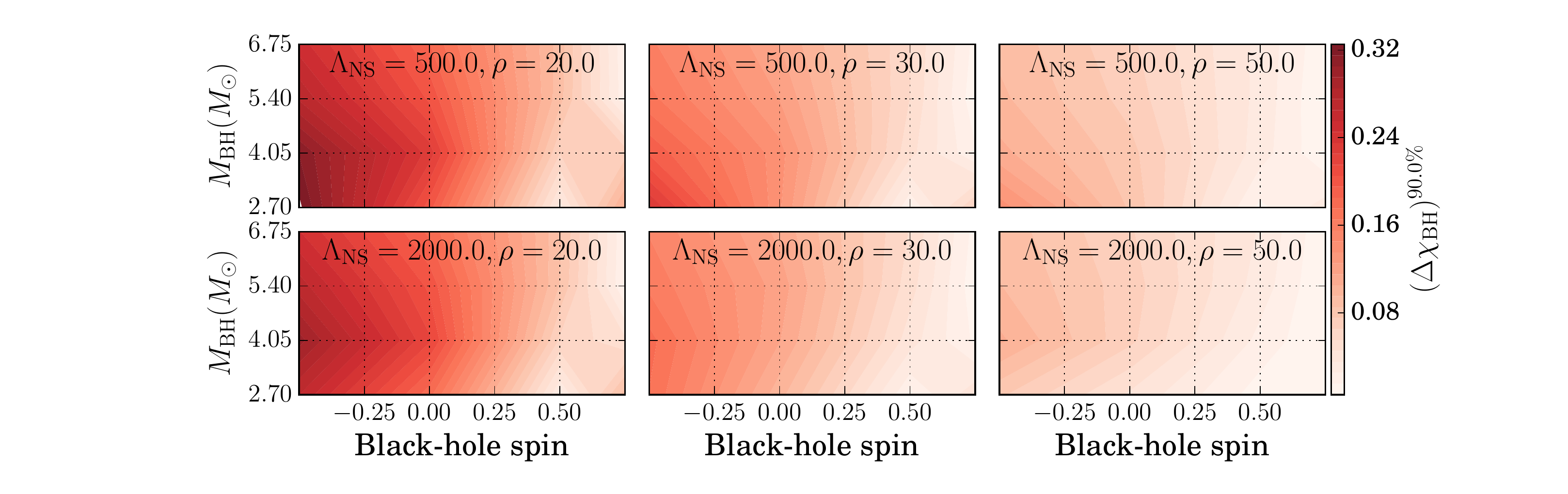}
\caption{{\bf Statistical measurement uncertainty for NSBH parameters,
ignoring tidal effects:}
We show here the statistical uncertainty associated with our measurement of
non-tidal parameters $\mchirp, \eta,$ and $\chibh$ (at $90\%$ credibility),
over the signal parameter space. Individual panels show the same as a function
of BH mass and spin. Across each row, we see the effect of increasing signal
strength (i.e. SNR) with the tidal deformability of the NS $\lambdans$ fixed.
Down each column, we see the effect of increasing $\lambdans$, at fixed SNR.
Tidal effects are ignored in templates.
}
\label{fig:CIWidths90_Lambda_SNR}
\end{figure*}
% #################
%%
In Fig.~\ref{fig:CIWidths90_Lambda_SNR}, we show how {\it precisely} can we
measure non-tidal NSBH parameters $X=\{\mchirp,\eta,\chibh\}$ using BBH templates.
The three panels correspond to $\mchirp$ (top), $\eta$ (middle), and $\chibh$
(bottom), and show the width of these credible intervals $(\Delta X)^{90\%}$
as a function of BH mass/spin (within each sub-panel), and NS properties, i.e.
$\lambdans$ (downwards in each column)~\footnote{We restrict NS mass to
$1.35M_\odot$ and its spin to zero. Varying its tidal deformability $\lambdans$
does not significantly change the measurement uncertainties for non-tidal
binary parameters, as is evident from comparing the two rows in each panel of
Fig.~\ref{fig:CIWidths90_Lambda_SNR}.}.
From the left-most column, we find that: (i) at $\rho=20$ the chirp mass is
measured remarkably well - to a precision of $0.16\%$ of its true value, and
(ii) so is $\chibh$. (iii) The dimensionless mass-ratio $\eta$ is determined
more loosely, with $25+\%$ uncertainty. If the signal is even louder
($\rho\geq 30$), all three measurements gain further precision, especially
$\eta$, for which the relative errors shrink down to single-digit percents.
% % 
We remind ourselves that these results do not tell the full story since the
precision of a measurement is only meaningful if the measurement is accurate 
to begin with. In our case there are tidal effects that have not been
incorporated into our search (BBH) templates, which can lead to a systematic
bias in parameter recovery. We refer the reader to Sec.~\ref{s1:PEwithnoNS} for
a comparative study of both systematic and statistical errors.

\section{Illustrations of Bayesian posteriors}\label{as1:illustrations}
\begin{figure*}
\centering
\includegraphics[width=1.05\columnwidth,trim=2cm 0 0 0]{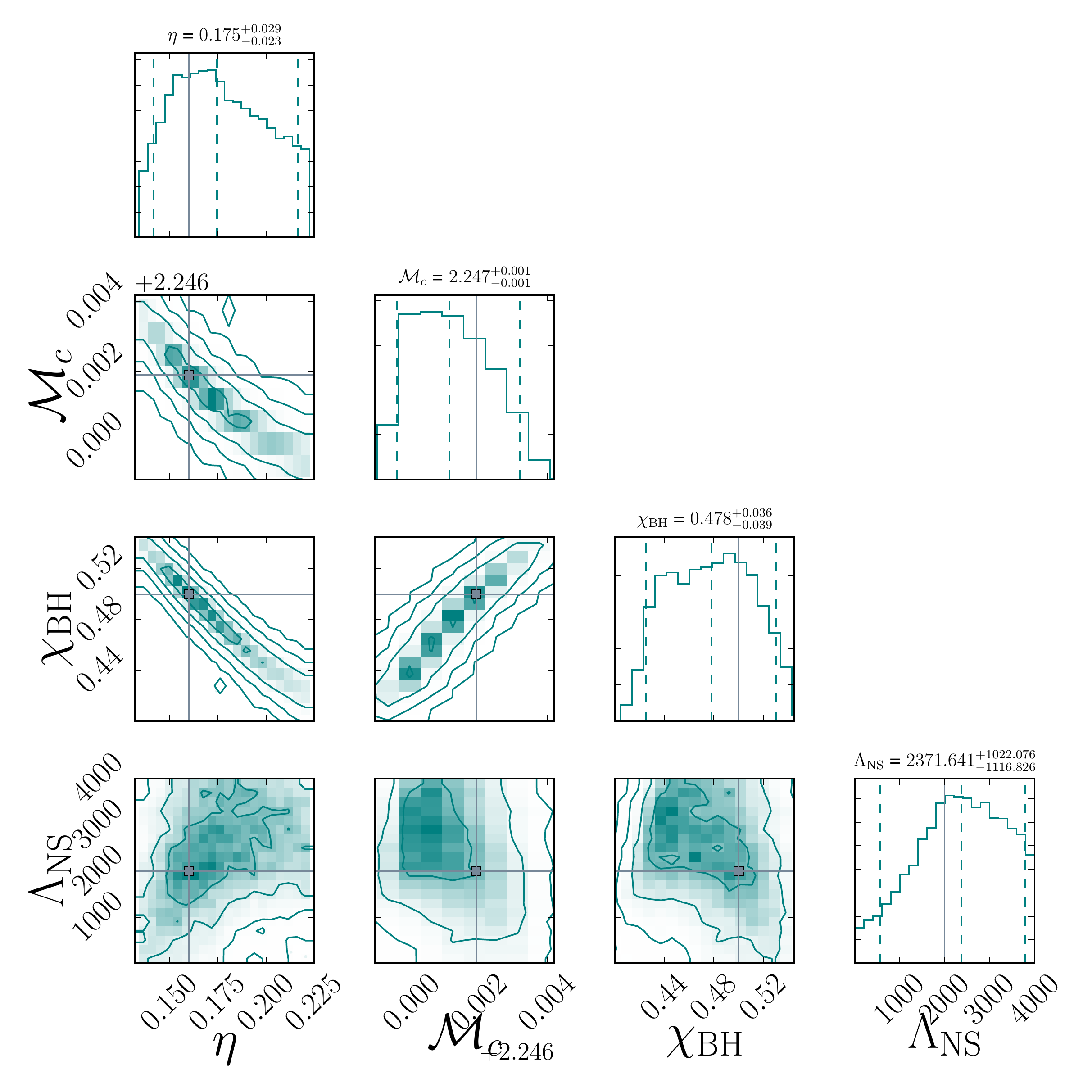}%\\
\includegraphics[width=1.05\columnwidth,trim=0.7cm 0 1cm 0]{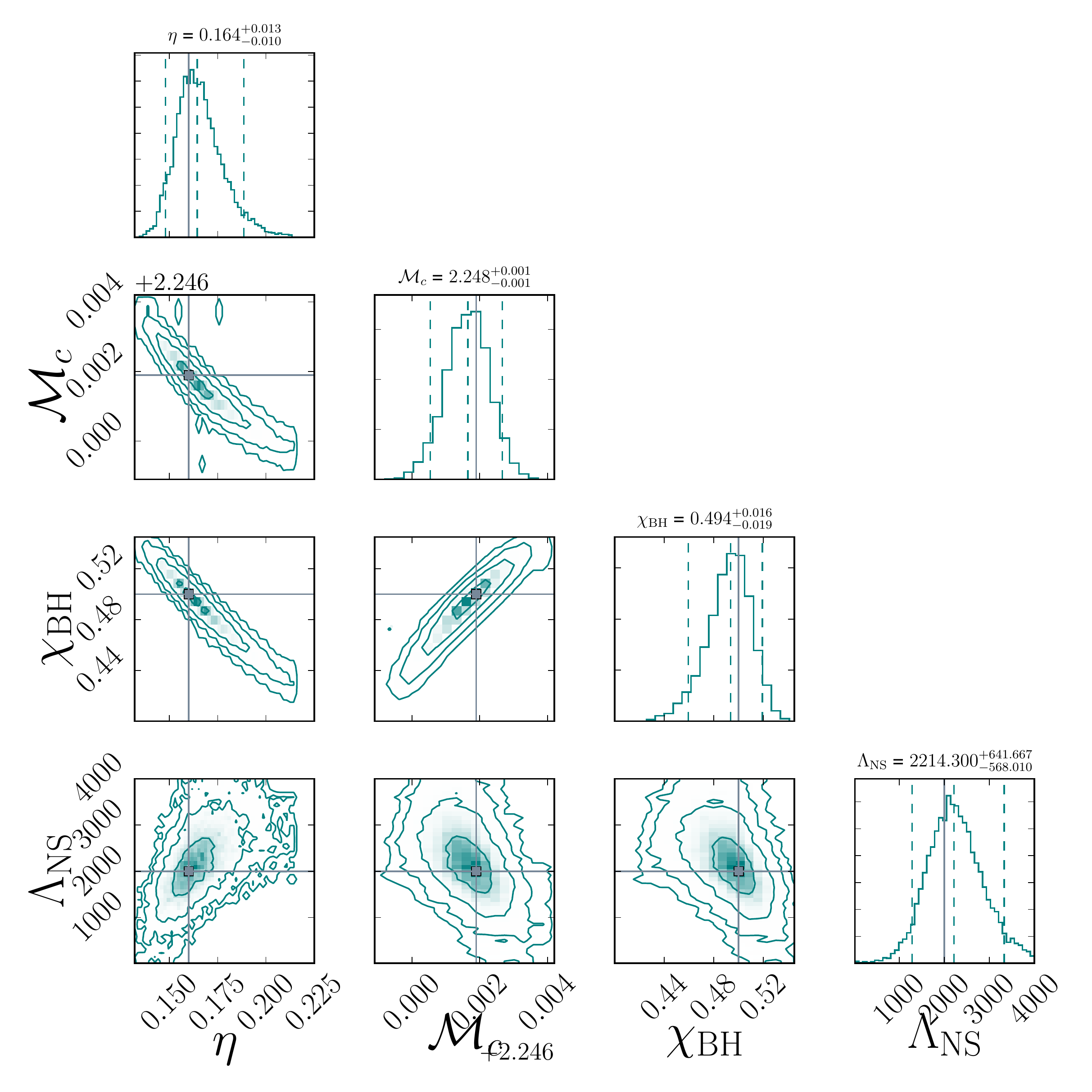}%\\
\caption{{\bf Illustrative posterior probability distributions for NSBH parameters,
for signals at different SNRs:}
We illustrate here two sets of two-dimensional joint probability distributions,
differing only in signal strength, with $\rho=20$ in the left panel, and
$\rho=50$ in the right. The injected parameters are 
$q = \mbh/\mns = 5.4M_\odot/1.35M_\odot = 4$, $\chibh=+0.5$, and 
$\lambdans=2000$. Contours are shown for $\{1-,2-,3-,\cdots\}\sigma$ confidence levels.
Templates include tidal effects, as evident in the bottom rows
of both panels which show the correlation of $\lambdans$ with non-tidal 
parameters. Contrasting the two panels illustrates the effect of increasing the
SNR on various parameter measurements.
}
\label{fig:SingleSystemLambda2DPDFs}
\end{figure*}
% #################

In Fig.~\ref{fig:SingleSystemLambda2DPDFs} we show the correlation of
mass, spin, and tidal parameter measurements. We keep the binary parameters
as in Fig.~\ref{fig:SingleSystemLambdaPDFvsSNR}, with $\lambdans=2000$, and
set $\rho=20$ (left panel) or $\rho=50$ (right panel).
We find that the measurement of $\lambdans$ is weakly degenerate with
other parameters, and at realistic SNRs it would improve by a few tens of 
percent if we knew non-tidal parameters to better accuracy. The predominant
factor that would enhance the measurement accuracy for $\lambdans$ is 
nevertheless the signal strength. Only when $\rho\gtrsim 50$ can we
expect $\lambdans$ measurement to be limited by its degeneracy with 
non-tidal parameters (at a factor of few level), as also reported by previous
studies~\cite{Lackey:2013axa}.

% ###########################################################
% ###########################################################
\section{Phenomenology of $\lambdans$ measurement errors}
% ###########################################################
% 
\begin{figure}
\centering    
\includegraphics[width=1.05\columnwidth]{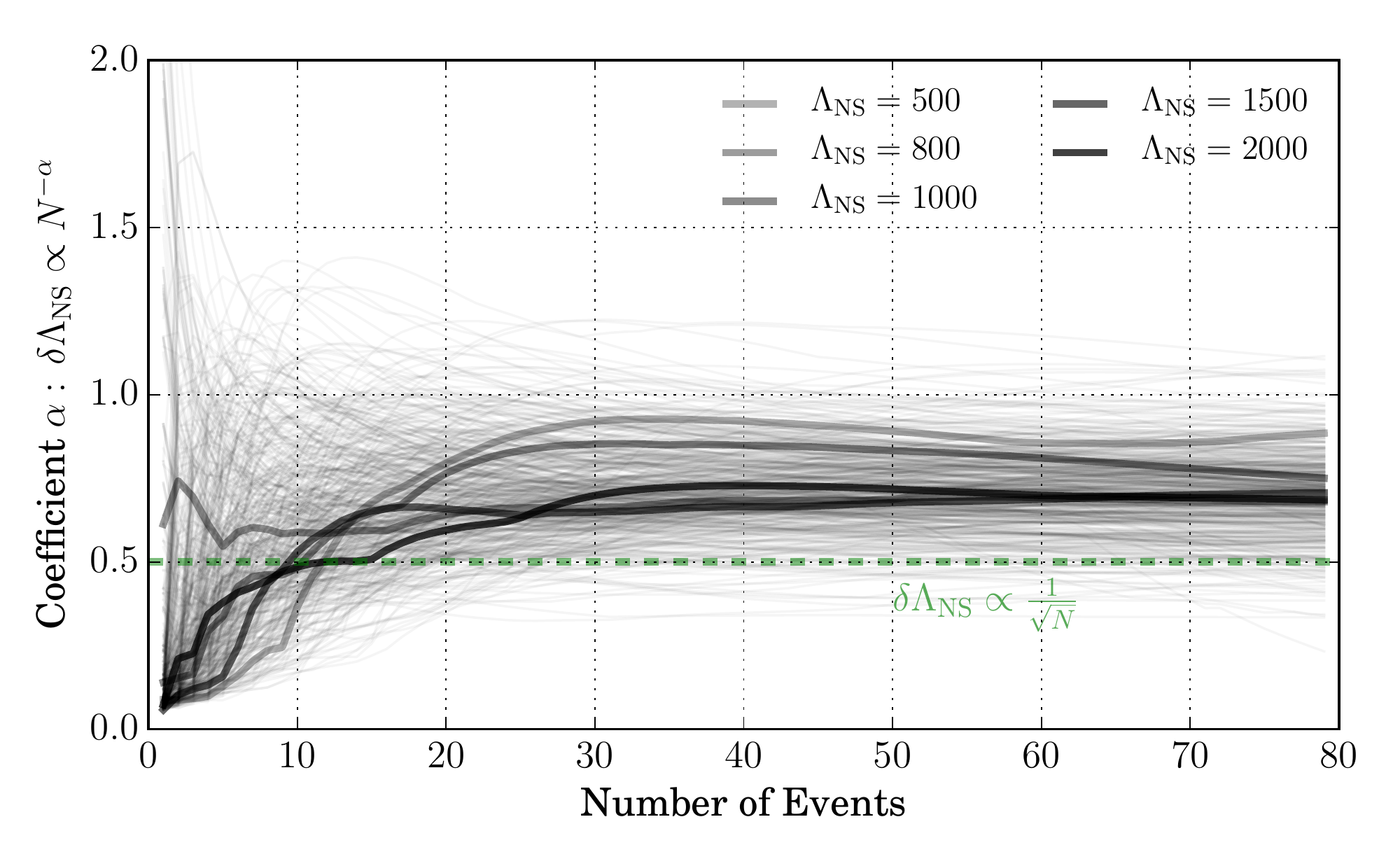}
\caption{%
Assuming a power-law dependence of the measurement error on the number of
events: $\delta\lambdans\propto 1/N^\alpha$, we show $\alpha$ in this figure
as a function of the number of observed events $N$. Shown are five families
of $100$ population draws each, with each family corresponding to one of
$\lambdans=\{500,800,1000,1500,2000\}$. Each grey curve corresponds to one
of these $100\times5 = 500$ populations. The thicker curves, one from each
family, shows the population we discussed in
Fig.~\ref{fig:TT_Lambda_vs_N_L800_CI90_0}-\ref{fig:TT_Lambda_vs_N_CI90_0}.
We find that a power-law is a good approximation for the concerned dependence,
and information accumulates {\it faster} than $1/\sqrt{N}$. We estimate
$\alpha\simeq 0.7^{+0.2}_{-0.2}$.
}
\label{fig:TT_PowerLawLambdaErrorVsN}
\end{figure}
\begin{figure}
\centering    
\includegraphics[width=\columnwidth]{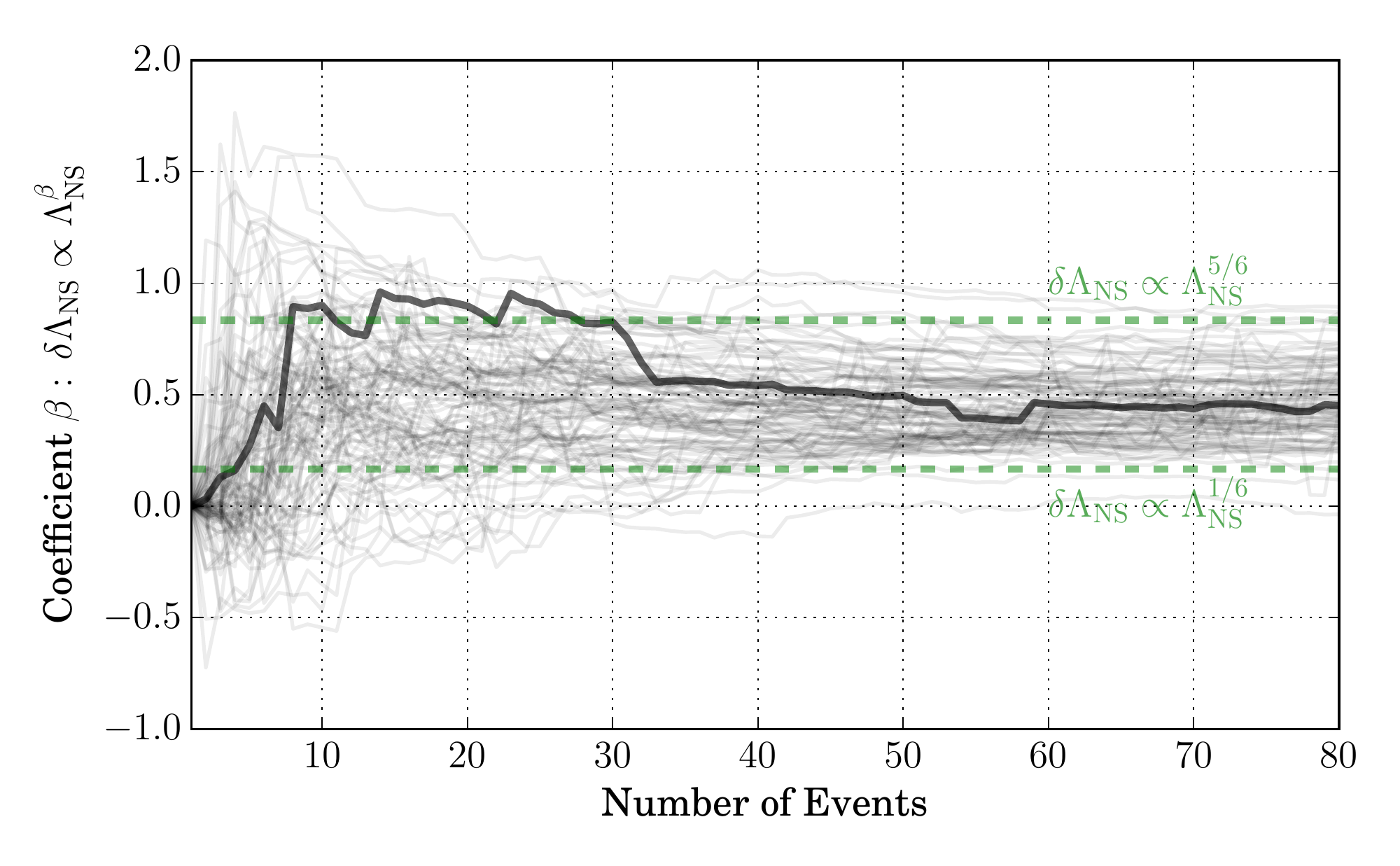}
\caption{%
In this figure, which is similar to Fig.~\ref{fig:TT_PowerLawLambdaErrorVsN},
we quantify the dependence of $\delta\lambdans$ on $\lambdans$ itself. Of 
the five families of simulated NSBH populations, we construct $100$
independent sets taking one population from each family. With each of 
these $100$ sets, and assuming a power-law dependence:
$\delta\lambdans\propto\lambdans^\beta$, we estimate $\beta$ and show it in
this figure as a function of the number of observed events $N$. The thicker
curve corresponds to the populations discussed in
Fig.~\ref{fig:TT_Lambda_vs_N_CI90_0}.
We find that $\beta$ can be estimated to lie within $[1/6,5/6]$ with a
likely value close to $1/2$. Since $0<\beta<1$, the relative error
$\delta\lambdans/\lambdans$ {\it decreases} as the star gets more 
deformable, while the absolute error $\delta\lambdans$ {\it increases}.
}
\label{fig:TT_PowerLawLambdaErrorVsLambda}
\end{figure}
Here, we quantitatively explore the dependence of our statistical
uncertainties for $\lambdans$ on the number of events, as well as on the true
NS deformability itself. First, we will focus on the dependence on $N$. We
assume a power-law dependence of the form
$\delta\lambdans\propto\ 1/N^\alpha$. For each of the $100$ populations 
for each of $\lambdans=500-2000$, we compute the exponent $\alpha$ as a
function of the number of observed events $N$, and show it in 
Fig.~\ref{fig:TT_PowerLawLambdaErrorVsN}. There are $100\times5=500$ curves
on the figure, with one highlighted for each value of population's $\lambdans$.
These highlighted values are only special in the sense that they correspond to
populations discussed earlier in this section (c.f.
Fig.~\ref{fig:TT_Lambda_vs_N_L800_CI90_0}-\ref{fig:TT_Lambda_vs_N_CI90_0}).
We immediately observe two things, (i) there is a globally similar dependence
on $N$ for all populations, and (ii) information accumulates {\it faster} than
$1/\sqrt{N}$. In fact, we find that if
$\delta\lambdans\propto\frac{1}{N^\alpha}$, $\alpha$ lines in the range
$0.7_{-0.2}^{+0.2}$.
Next, we focus on the dependence of $\delta\lambdans$ on $\lambdans$ of the
population itself. As suggested by Fisher-matrix studies~\cite{Lackey:2013axa},
and as for $N$, we assume the form $\delta\lambdans\propto\lambdans^\beta$.
From each set of $100$ populations with a given $\lambdans$ value, we draw one
at random, and form a set of $5$ similarly drawn populations, one for each of
$\lambdans=\{500,800,1000,1500,2000\}$. With each set, we determine $\beta$
for different number of observed events $N$. In all, we make $100$ independent
$5-$population sets and show the value of $\beta$ measured from each in 
Fig.~\ref{fig:TT_PowerLawLambdaErrorVsLambda}. We find that the assumed
relation $\delta\lambdans\propto\lambdans^\beta$ gets fairly robust for 
larger values of $N$, with $\beta$ converging to $\beta=0.5^{+0.33}_{-0.33}$.
The fact that $0<\beta<1$ implies that the relative error
$\delta\lambdans/\lambdans$ {\it decreases} with increasing $\lambdans$, while
the absolute error {\it increases}.
From these results, we conclude that the measurement uncertainty for
$\lambdans$ after $N$ observations is
\begin{equation}
 \delta\lambdans\propto \dfrac{\lambdans^{0.5^{+0.33}_{-0.33}}}{N^{0.7_{-0.2}^{+0.2}}}.
\end{equation}
We also find that while these results are inferred from paradigm A populations,
paradigm B gives very similar results.

% ###########################################################
% ###########################################################
\section{Choice of underlying BBH model in LEA}
% ###########################################################
% 
\begin{figure}
\centering    
\includegraphics[width=1.05\columnwidth]{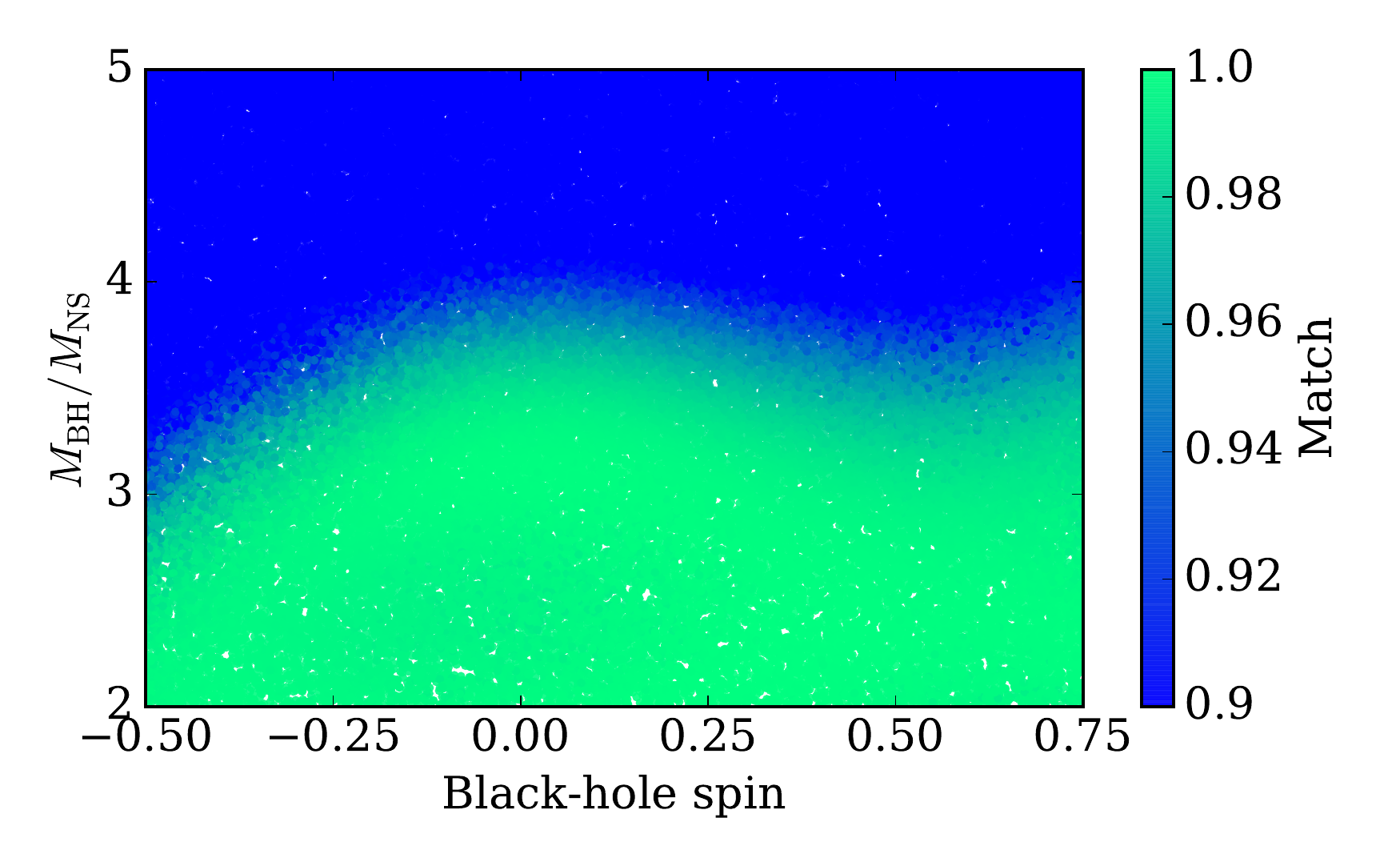}
\caption{%
We compare two alternatives of the tidal NSBH model from Ref.~\cite{
Lackey:2013axa}, which differ in their underlying BBH prescriptions. One which
we use in this study uses SEOBNRv2, while the other uses IMRPhenomC as its
base. In this figure, we show the normalized overlap (match) between the two for
$2,000,000$ points sampled uniformly in the NSBH parameter space. We find that
for $q\gtrsim 4$ the discrepancies between IMRPhenomC and SEOBNRv2 as reported
in~\cite{Kumar:2015tha} dominate over tidal terms.
\label{fig:PhenomC_vs_SEOBNRv2_LEA}
}
\end{figure}
The waveform model used in this paper is a variant of those calibrated
in Ref~\cite{Lackey:2013axa}. In that work, the authors also calibrate
a tidal prescription with the phenomenological model IMRPhenomC~\cite{
Santamaria:2010yb}
as the base BBH model. Previous work~\cite{Kumar:2015tha,Kumar:2016dhh} has 
shown that IMRPhenomC can exhibit pathological behavior for mass-ratios
$q\gtrsim 4$ and/or non-zero black hole spins. We compute noise-weighted
inner-products between the two variants for $2,000,000$ points sampled
over the NSBH parameter space, and show the results in 
Fig.~\ref{fig:PhenomC_vs_SEOBNRv2_LEA}. We restrict the comparison to frequencies
that are affected by the tidal disruption of the NS, by integrating
the inner-products from $f = \mathrm{max}(15, 0.01/M)$~Hz
(where $M$ is expressed in seconds ($1M_\odot \simeq 4.925\mu$S, see
Eq.~(32-34) of~\cite{Lackey:2013axa}).
We find that the differences between the two variants of LEA+ have mismatches
of a few percent, while the tidal corrections contribute at a sub-percent
level. We conclude that the differences of the underlying BBH model in LEA+
dominate over its tidal calibration, and since SEOBNRv2 has been shown to
be more reliable than IMRPhenomC~\cite{Kumar:2015tha,Kumar:2016dhh}, we 
recommend the use of SEOBNRv2-based LEA+ in upcoming LIGO-Virgo analyses.

\end{appendix}
%%%%%%%%%%%%%%%%%%%%%%%%%%%%%%%%%%%%%%%%%%%%%%%%%%%%%%%%%%%%%%%%%%%%%%%%%%%%%%%

%%%%%%%%%%%%%%%%%%%%%%%%%%%%%%%%%%%%%%%%%%%%%%%%%%%%%%%%%%%%%%%%%%%%%%%%%%%%%%%
\section*{References}
%%%%%%%%%%%%%%%%%%%%%%%%%%%%%%%%%%%%%%%%%%%%%%%%%%%%%%%%%%%%%%%%%%%%%%%%%%%%%%%
\bibliography{paper}

\end{document}